\documentclass[12pt]{article}
\usepackage{graphicx}

\begin{document}

\begin{center}

\baselineskip 30pt

\vskip 2cm

{\Large {\bf An ansatz to the quantum phase transition in a
dissipative two-qubit system}}

\vskip 1cm

{\large Hang Zheng$^1$, Zhiguo L\"{u}$^1$,  and Yang Zhao$^2$}

{$^1$Key Laboratory of Artificial Structures and Quantum Control
(Ministry of Education), Department of Physics, Shanghai Jiao Tong
University, Shanghai 200240, China}

{$^2$Division of Materials Science, Nanyang
Technological University, Singapore 639798, Singapore}

\vskip 1cm

{\bf Abstract}

\end{center}

\baselineskip 20pt

By means of a unitary transformation, we propose an ansatz to study
quantum phase transitions in the ground state of a two-qubit system
interacting with a dissipative reservoir. First, the ground state
phase diagram is analyzed in the presence of the Ohmic and sub-Ohmic
bath using an analytic ground state wave function which takes into
account the competition between intrasite tunneling and intersite
correlation. The quantum critical point is determined as the
transition point from non-degenerate to degenerate ground state and
our calculated critical coupling strength $\alpha_c$ agrees with
that from the numerical renormalization group method. Moreover, by
computing the entanglement entropy between the qubits and the bath
as well as the qubit-qubit correlation function
in the ground state, we explore the nature of the quantum phase
transition between the delocalized and localized states.

\baselineskip 20pt

\vskip 1cm


\pagebreak

\baselineskip 20pt

\section{Introduction}

Quantum phase transitions (QPT) in impurity models with
competing interactions have been a subject of great interest in
recent years. In this work we consider a two-qubit system
coupled with a dissipative bath, in which the competing
interactions are the intrasite tunneling, the qubit-bath coupling,
and the intersite qubit-qubit interaction. The Hamiltonian for the
interacting system and environment reads\cite{hur}
\begin{eqnarray}
&&H=\sum_{i=1,2}\left\{-\frac{\Delta}{2}\sigma^x_i-\frac{\epsilon}{2}\sigma^z_i+\sum_k
\frac{g_k}{2}(b^{\dag}_k+b_k)\sigma^z_i\right\} +K \sigma^z_1
\sigma^z_2+\sum_k\omega_k b^{\dag}_k b_k.
\end{eqnarray}
where $b^{\dag}_k$ ($b_k$) is the creation (annihilation) operator
of boson mode with frequency $\omega_k$ and $\sigma^x$ and
$\sigma^z$ are the Pauli matrices where the subscripts denote qubit
1 and 2. $\Delta$ is the intrasite tunneling, $\epsilon$ is the bias
on every qubit, and $K$ is the Ising-type qubit-qubit interaction. Throughout this paper we set $\hbar=1$.
The qubit-bath coupling is denoted by $g_k$, and the effect of the bath is
characterized by a spectral density
$J(\omega)=\sum_kg^2_k\delta(\omega-\omega_k)=2\alpha
\omega^s\omega^{1-s}_c\theta(\omega_c-\omega)$ with the
dimensionless coupling strength $\alpha$ and the hard upper cutoff
at $\omega_c$. The index $s$ accounts for various physical
situations\cite{rmp,book}: the Ohmic $s=1$, sub-Ohmic $s<1$ and
super-Ohmic $s>1$ baths. In this paper we use a very small bias
$\epsilon/\omega_c\le 10^{-5}$ to trigger the QPT\cite{hur}.

The QPT is a ground state transition when the parameter of
Hamiltonian changes across some critical point. If the qubits and
bath are decoupled, $g_k=0$, Hamiltonian (1) can be solved easily
and there is no QPT if we keep a very small bias
$\epsilon/\omega_c\le 10^{-5}$. The QPT is triggered by competing
interactions: The intrasite tunneling $\Delta$ favors the
delocalized state with $\langle\sigma^z_i\rangle_G\approx 0$, where $i=1,2$ and
$\langle ... \rangle_G$ denotes the ground state average.
But the role of a finite qubit-bath coupling strength ($g_k\neq 0$, or finite $\alpha$) is to
ensure dissipation in the qubits\cite{rmp,book}, which competes with the tunneling effect and
leads to the possibility of localization with a finite value of
$\langle\sigma^z_i\rangle_G$. The QPT in the
single-qubit spin-boson model (SBM) was studied by many authors and
its properties are well-understood. Various numerical methods were
used for this purpose, such as the numerical renormalization group
(NRG)\cite{bu1,bu2,hur1}, the quantum Monte Carlo (QMC)\cite{qmc},
the method of sparse polynomial space representation\cite{fehske},
the extended coherent state approach\cite{chen}, and the variational
matrix product state approach\cite{vmps}. In addition, an
extension of the Silbey-Harris \cite{sh} ground state has been recently employed by
us \cite{zheng1} to study the QPT of the single-qubit SBM in the
Ohmic ($s=1$) and sub-Ohmic ($s<1$) bath.

For the two-qubit SBM described by Eq.~(1) where the qubits interact
with a common bath, the QPT may differ significantly from that of
the single-qubit SBM because the qubit-bath interaction may induce
an effective Ising-type ferromagnetic coupling between qubits which
is superposed on the original Ising coupling $K$ and leads to a
renormalized Ising coupling $(K-V)\sigma^z_1\sigma^z_2$, where $-V$
is the induced coupling strength \cite{hur}. For the two-qubit SBM
with the Ohmic bath ($s=1$), McCutcheon {\it et al.} predicted
variationally the quantum critical point (QCP) at $\alpha_c=0.5$ in
the absence of both bias ($\epsilon=0$) and direct Ising couple
($K=0$) \cite{nazir}. Using the numerical renormalization group,
however, Orth {\it et al.}\cite{hur} arrived at $\alpha_c\approx
0.15$.
Recently, Winter and Rieger studied the quantum phase transition of multi-qubit SBM for $K=0$
with the help of extensive quantum Monte Carlo simulations\cite{winter}.
They found $\alpha_c \approx 0.2$ for $\Delta/\omega_c=0.1$ in the Ohmic bath.

In this work we present a new analytical approach based on a
unitary transformation. We will show that due to the
renormalized Ising coupling the QCP of the two-qubit SBM acquires a substantial
shift as compared to that of the single-qubit case. In addition, the
qubit-bath entanglement entropy will be calculated to see how the
parameters in (1), $\Delta$, $\alpha$, and $K$, compete with each
other and lead to the delocalization-localization transition.

The remainder of the paper is organized as follows. In Section II,
the unitary transformation of the Hamiltonian is introduced, and the
ground state properties are discussed. Implications of our results
to the quantum phase transition are elaborated in  Section III. The
entanglement entropy between the qubits and the bath and the qubit-qubit correlation
function are studied in sections IV and V, respectively. Finally, conclusions are drawn in Section VI.

\section{Unitary transformation}

In order to find the ground state, we apply a unitary transformation
on Hamiltonian (1), i.e., $H^\prime=\exp (S)H \exp(-S)$, with the
generator S given by
\begin{eqnarray}
&&S=\sum_{\bf k}\frac{g_{\bf k}}
    {2\omega_{\bf k}}(b^{\dag}_k-b_k)
    \left [ \xi_k(\sigma^z_1+\sigma^z_2)+(1-\xi_k)\sigma_0\right ].
\end{eqnarray}
where $\sigma_0$ is a number and $\xi_k$ is a function of
$\omega_k$. Compared with the ground state of Ref.\cite{nazir}, a
finite number $\sigma_0$ is introduced to take into account the
modified bias $\epsilon\to \epsilon'$ (when $\epsilon\neq 0$)
because of the qubit-bath interaction\cite{zheng2,chin,NJ}. The form
of $\sigma_0$ and $\xi_k$ will be determined later. After the
transformation, we obtain
\begin{eqnarray}
&&H'=H'_0+U_{\epsilon}+H'_1+H'_2 ,\\
&&H'_0=-\eta\Delta(\sigma^x_1+\sigma^x_2)/2+(K-V)\sigma^z_1\sigma^z_2
 +\sum_k\omega_k b^{\dag}_k b_k-V+F\sigma^2_0/4 ,\\
&&U_{\epsilon}=-\epsilon'(\sigma^z_1+\sigma^z_2)/2,~~~\epsilon'=\epsilon+F\sigma_0\\
&&H'_1=\sum_k g_k(b^{\dag}_k+b_k)(1-\xi_k)( \sigma^z_1+\sigma^z_2-\sigma_0 )/2 \nonumber\\
&&-\eta\Delta\sum_k\frac{g_k}{2\omega_k}\xi_k(b^{\dag}_k-b_k)
    ( i\sigma^y_1 +i\sigma^y_2 ),\\
&&H'_2=-\frac{\Delta}{2}(\sigma^x_1+\sigma^x_2)\left \{
\cosh(Y)-\eta \right\}-\frac{\Delta}{2}
(i\sigma^y_1+i\sigma^y_2)\left\{\sinh(Y)- \eta Y\right\},
\end{eqnarray}
where $F=\sum_k g^2_k(1-\xi_k)^2/\omega_k$ and $Y=\sum_k
g_k\xi_k(b^{\dag}_k-b_k)/\omega_k$. In the zeroth-order transformed
Hamiltonian $H'_0$,
\begin{eqnarray}
&&\eta=\exp\left\{-\sum_k \frac{g^2_k}
     {2\omega^2_k}\xi^2_k\right\}
\end{eqnarray}
is the environment dressing of the bare tunneling $\Delta$, and
\begin{eqnarray}
&&V=\sum_k \frac{g^2_k}{2\omega_k}\xi_k(2-\xi_k)
\end{eqnarray}
is the bath induced Ising-type interaction. Note that in $H'_0$ the
Ising-type interaction is modified by the qubit-bath coupling:
$K^\prime = K-V$. Besides, $\epsilon'$ in Eq.(5) is the modified
bias which is related to the number $\sigma_0$ introduced in our
transformation.

With only the Ising-type interaction, the zeroth-order Hamiltonian
$H'_0$ may be diagonalized by the following two-qubit states,
\begin{eqnarray}
&&|A\rangle=\left [(u+v)|11\rangle+(u-v)|22\rangle\right ]/\sqrt{2},\\
&&|B\rangle=\left [|12\rangle+|21\rangle\right ]/\sqrt{2},\\
&&|C\rangle=\left [|12\rangle-|21\rangle\right ]/\sqrt{2},\\
&&|D\rangle=\left [(v-u)|11\rangle+(v+u)|22\rangle\right ]/\sqrt{2},
\end{eqnarray}
where $|1\rangle$ and $|2\rangle$ are eigenstates of $\sigma^x$:
$\sigma^x|1\rangle=|1\rangle$ and $\sigma^x|2\rangle=-|2\rangle$,
and $|12\rangle$ denotes that the state of first qubit is
$|1\rangle$ and that of second one is $|2\rangle$. The parameters
$u$ and $v$ are given by
\begin{eqnarray}
&&u=\frac{1}{\sqrt{2}}\sqrt{1+(V-K)/W},\mbox{~~~}
v=\frac{1}{\sqrt{2}}\sqrt{1-(V-K)/W},
\end{eqnarray}
where $W=\sqrt{\eta^2\Delta^2+(V-K)^2}$. Thus, the qubit dependent
part of $H'_0$ may be diagonalized as
\begin{eqnarray}
&&H'_0=-W\left(|A\rangle\langle A|-|D\rangle\langle
D|\right)-(V-K)\left(|B\rangle\langle B|-|C\rangle\langle C|\right)
 \nonumber\\
&& +\sum_k\omega_k b^{\dag}_k b_k-V+F\sigma^2_0/4 ,
\end{eqnarray}
and $U_{\epsilon}$ in Eq.~(5) becomes
\begin{eqnarray}
&&U_{\epsilon}=-\epsilon'\left\{(u|A\rangle+v|D\rangle)\langle
B|+|B\rangle(u\langle A|+v\langle D|)\right\}.
\end{eqnarray}
In this work we consider only the case of weak bias with
$\epsilon/\omega_c\le 10^{-5}$\cite{hur}. At the lowest order
of $\epsilon$ we can diagonalize $H'_0+U_{\epsilon}$ in the space
expanded by $|A\rangle$ and $|B\rangle$,
\begin{eqnarray}
&&|A\rangle=\cos\theta|G\rangle-\sin\theta|X\rangle,~
|B\rangle=\sin\theta|G\rangle+\cos\theta|X\rangle,
\end{eqnarray}
where
\begin{eqnarray}
&&\cos\theta=\frac{1}{\sqrt{2}}\left(1+\frac{W-V+K}{\Sigma}\right)^{1/2},~
\sin\theta=\frac{1}{\sqrt{2}}\left(1-\frac{W-V+K}{\Sigma}\right)^{1/2},\\
&&\Sigma=\sqrt{(W-V+K)^2+4\epsilon^{\prime 2}u^2}.\nonumber
\end{eqnarray}
Then we have
\begin{eqnarray}
&&H'_0+U_{\epsilon}=-{1\over 2}[W+V-K+\Sigma]|G\rangle\langle
G|-{1\over 2}[W+V-K-\Sigma]|X\rangle\langle
X|\nonumber\\
&&+(V-K)|C\rangle\langle C|+W|D\rangle\langle D|+\sum_k\omega_k
b^{\dag}_k b_k-V+F\sigma^2_0/4
 \nonumber\\
&&
-\epsilon'v\left\{(-\sin\theta|G\rangle+\cos\theta|X\rangle)\langle
D|+h.c.\right\} .
\end{eqnarray}
It is easy to see that if the last term in
Eq.~(19) is neglected, the ground state of $H'_0+U_{\epsilon}$ is $|G\rangle$, and
in this work we are mainly concerned with the ground state
properties. In Eq.~(19), the coefficient of the
transition term $|G\rangle\langle D|+|D\rangle\langle G|$ is
$\epsilon'v\sin\theta\propto \epsilon'^2$. In numerical
calculations we use a very small bias $\epsilon/\omega_c\le 10^{-5}$
to trigger the QPT \cite{hur} while staying in the range of
$\epsilon'/\omega_c\le 0.05$, and consequently, the transition term
$|G\rangle\langle D|+|D\rangle\langle G|$ can be dropped safely.
Fortunately, the QCP at $\alpha\sim\alpha_c$ falls within this range, and
our numerical calculations are carried out in the range of
$0\le\alpha\le 1.1\alpha_c$.

The first-order Hamiltonian $H'_1$ can be recast as
\begin{eqnarray}
&&H'_1=\sum_k g_k b^{\dag}_k
    \bigg \{ (1-\xi_k)\left
    [u(\cos\theta|G\rangle-\sin\theta|X\rangle)(\sin\theta\langle G|+\cos\theta\langle X|)
+h.c.-\frac{\sigma_0}{2}\right ]\nonumber\\
&&+\left.\frac{\eta\Delta}{\omega_k}\xi_k \left
[v(\cos\theta|G\rangle-\sin\theta|X\rangle)(\sin\theta\langle
G|+\cos\theta\langle X|)-h.c.\right ]\right\}+h.c. \nonumber\\
&&=\sum_k g_k (b^{\dag}_k+b_k)
    (1-\xi_k)\left
    [u\sin(2\theta)(|G\rangle\langle G|-|X\rangle\langle X|)-\frac{\sigma_0}{2}\right ]\\
&&+\sum_k g_k b^{\dag}_k
    \left
    [u(1-\xi_k)\cos(2\theta)(|G\rangle\langle X|+|X\rangle\langle G|)+v\frac{\eta\Delta}{\omega_k}\xi_k(|G\rangle\langle X|-|X\rangle\langle G|)
    \right ]+h.c.,\nonumber
\end{eqnarray}
where $h.c.$ is short for the Hermitian conjugate. Then, if we choose
\begin{eqnarray}
&&\sigma_0=2u\sin(2\theta)=\frac{4u^2\epsilon'}{\Sigma},~~~
\xi_k=\frac{\omega_k}{\omega_k+\Sigma},
\end{eqnarray}
we have $H'_1|G\rangle |\{0_k\}\rangle=0$, where $|\{0_k\}\rangle$
is the vacuum state of the environment. Now we can see clearly the
reason why we introduce the term $(1-\xi_k)\sigma_0$ in Eq.(2) for
the generator $S$. Note that the term $\xi_k(\sigma^z_1+\sigma^z_2)$
in $S$ comes from the Silbey-Harris type ansatz where
$\xi_k=\omega_k/(\omega_k+\Sigma)\approx 1$ for the high-frequency
oscillators. However, $1-\xi_k=\Sigma/(\omega_k+\Sigma)\approx 1$
for the lower-frequency oscillators and this is to say that when
$\sigma_0\neq 0$ the lower-frequency oscillators may play an
important role. We will see in next section that away from the QPT
($\alpha<\alpha_c$) we have $\sigma_0\approx 0$ and the dynamic
displacement in $S$, $\xi_k(\sigma^z_1+\sigma^z_2)$, dominates; but
around the QCP $\alpha\sim\alpha_c$, $\sigma_0\neq 0$ and the static
displacement $(1-\xi_k)\sigma_0$ comes into play.

Since $H'_1|G\rangle |\{0_k\}\rangle=0$, the ground state of
$H'_0+U_{\epsilon}+H'_1$ is $|G\rangle |\{0_k\}\rangle$ with the
ground state energy,
\begin{eqnarray}
&&E_g=-{1\over 2}[W+V-K+\Sigma]-V+\sum_k\frac{g^2_k}
     {4\omega_k}(1-\xi_k)^2\sigma^2_0.
\end {eqnarray}
This ground state energy can also be derived from the variational
principle. Our theory is to introduce a trial ground state of the
original Hamiltonian $H$ (Eq.(1)),
\begin{eqnarray} \label{gsf}
&&|g.s.\rangle=\exp(-S)|G\rangle |\{0_k\}\rangle.
\end{eqnarray}
The ground state energy is Eq.(22): $$E_g=\langle g.s.
|H|g.s.\rangle=\langle\{0_k\}|\langle G|\exp(S)H\exp(-S)|G\rangle
|\{0_k\}\rangle$$ (Note that $\langle\{0_k\}|\langle G|H'_2|G\rangle
|\{0_k\}\rangle=0$). If $\sigma_0=0$, our ground state is the same
as the variational ground state of Ref.\cite{nazir}. But for
$\alpha\ge\alpha_c$, we introduce a finite $\sigma_0$ which can be
determined by the ground state variation: $\partial
E_g/\partial\sigma_0=0$. It is easily to prove that $\partial
E_g/\partial\sigma_0=0$ leads to Eq.(21) for $\sigma_0$. We will
show in next section that a nonzero $\sigma_0$ leads to a nonzero
$\langle\sigma_z\rangle\neq 0$ which determines the QCP.

Furthermore, the ground state average of $\sigma^x$ is
\begin{eqnarray}
&&\langle\sigma^x_1\rangle_G=\langle\sigma^x_2\rangle_G= {1\over
2}\langle
g.s.|(\sigma^x_1+\sigma^x_2)|g.s.\rangle=\frac{\eta^2\Delta}{W}\cos^2\theta.
\end{eqnarray}
The numerical results of $E_g$ and $\langle\sigma^x\rangle_G$ will
be shown in next section.

\section{Quantum phase transition}

We use the same criterion as that used in Ref.\cite{hur} to
determine the critical coupling in this work, that is, the emergence
of a non-zero ground state expectation of $\langle\sigma^z\rangle$
as the coupling $\alpha$ increases across some critical point
$\alpha_c$. We note that this criterion is  different from that of
Ref.\cite{nazir}, where the vanishing of the renormalized tunneling
$\eta\to 0$ is used as the criterion. Since
$\epsilon'=\epsilon+F\sigma_0$, Eq.~(21) leads to
\begin{eqnarray}
&&\sigma_0=\frac{4u^2\epsilon}{\Sigma}\left/\left(1-\frac{4u^2
F}{\Sigma}\right)\right..
\end{eqnarray}
The ground state average of $\sigma^z$ is
\begin{eqnarray}\label{sz}
&&\langle\sigma^z_1\rangle_G=\langle\sigma^z_2\rangle_G= {1\over
2}\langle G|(\sigma^z_1+\sigma^z_2)|G\rangle=u\sin(2\theta)
=\frac{2\epsilon'u^2}{\Sigma}=\frac{\sigma_0}{2}.
\end{eqnarray}
As $\epsilon/\omega_c<10^{-5}$ is very small, Eq.(18) leads to
$\Sigma\approx W-V+K$ for the delocalized phase. In this phase
$\sigma_0\sim\epsilon$ is also very small until
\begin{eqnarray}
&&1-\frac{4u^2 F}{W-V+K}=0,
\end{eqnarray}
where a quantum phase transition occurs, and a finite average
$\langle\sigma^z_1\rangle=\langle\sigma^z_2\rangle$ emerges. That
is, the two-qubit SBM exhibits two ground state phases \cite{hur}: a
delocalized phase in which $\langle\sigma^z_{1,2}\rangle\to 0$ in
the limit of $\epsilon\to 0$, and a localized phase with
$\langle\sigma^z_{1,2}\rangle\neq 0$ even in the presence of an
infinitesimal bias $\epsilon =0^{+}$. Note that $\sigma_0$ (Eq.(25))
is not divergent at the transition point and in the localized phase
because $1-4u^2F/\Sigma>0$ ($\Sigma$ is defined in Eq.(18)) and
$\epsilon'>0$ even if $\epsilon=0^+$.

The critical coupling strength at the QCP $\alpha_c$ can be
determined by Eq.~(27) because $F\propto\alpha$. For effectively
ferromagnetic coupling ($K-V<0$) in the zeroth-order Hamiltonian
$H'_0$ of Eq.~(4), it is found that $W-V+K\approx
0.5\eta^2\Delta^2/(V-K)$ in the scaling limit of $\Delta\ll
\omega_c$, and to the lowest order of $\Delta/\omega_c$, we have
\begin{eqnarray}
&&F=2\alpha\omega^{1-s}_c\int^{\omega_c}_0\frac{(W-V+K)^2\omega^{s-1}d\omega}{(\omega+W-V+K)^2}
\sim\frac{2\pi\alpha\omega_c(1-s)}{\sin[\pi(1-s)]}\left\{\frac{W-V+K}{\omega_c}\right\}^s.
\end{eqnarray}
Then, Eq.~(27) becomes
\begin{eqnarray}
&&1-\frac{4\pi\alpha_c(1-s)(W+V-K)}{\sin[\pi(1-s)]W}\left(\frac{W-V+K}{\omega_c}\right)^{s-1}=0.
\end{eqnarray}
When $K<V$ and $\Delta\ll \omega_c$, $W-V+K\approx
0.5\eta^2\Delta^2/(V-K)$ and $(W+V-K)/W\approx 2$. Then, it is
easily seen that $\alpha_c=1/8+O(\Delta/\omega_c)$ for $s=1$, and
$\alpha_c=0+O(\Delta/\omega_c)$ for $s<1$. In the super-Ohmic regime of $s>1$,
$\alpha_c\to\infty$, and the system is always in the delocalized
state in the limit of $\epsilon\to 0$. Our estimation is comparable to
those of Ref.~\cite{hur}: $\alpha_c=0.15+\mathcal{O}(\Delta/\omega_c)$ for
$s=1$ and $\alpha_c=0+\mathcal{O}(\Delta/\omega_c)$ for $s<1$. Moreover, it is also interesting to list the
prediction of Ref.~\cite{nazir}: $\alpha_c=0.5$ for $s=1$.

For finite values of $\Delta/\omega_c$, the QCP can be determined by
Eq.~(27). Figure~1 is the $\alpha$-versus-$\Delta$ phase diagram for
various values of $s$ with $K=0$ and very weak bias
$\epsilon/\omega_c=10^{-5}$. One can see that in the scaling limit
$\Delta/\omega_c\to 0$, $\alpha_c\to 0.125$ for the Ohmic bath
$s=1$, and $\alpha_c\to 0$ for the sub-Ohmic bath $s<1$. Meanwhile, $\alpha_c$
increases with increasing tunneling $\Delta$ because a larger
tunneling strength favors the delocalized state.

Figure~2 is the $\alpha$-versus-$K$ phase diagram for various values
of $s$ with $\Delta/\omega_c=0.1$ and very weak bias
$\epsilon/\omega_c=10^{-5}$, which is similar to Figs.~2 and 3 in
Ref.\cite{hur}. As the effective Ising interaction in $H'_0$ is
$(K-V)\sigma^z_1\sigma^z_2$, a positive (antiferromagnetic) $K$
reduces the bath-induced interaction $-V$, while a negative
(ferromagnetic) $K$ enhances it. This explains that in the phase
diagram a positive $K$ favors the delocalized phase while a negative
$K$ is unfavorable to it. One can see that the phase boundary
depends on K very weakly for the ferromagnetic case ($K<0$), while for
the antiferromagnetic case ($K>0$) the delocalized region extends to
a larger $\alpha_c$, and the asymptotic line of the phase boundary
for a larger $K>0$ is given by $K_r=K-\alpha\Omega_c/s=0$ ($K_r$ is the
renormalized Ising coupling defined by Ref.~\cite{hur}).
We present a comparison of the NRG results and ours in Figs.~2(b) (Ohmic bath) and 2(c) (sub-Ohmic bath). For $K<0$, the phase boundary of $\alpha_c$ for the Ohmic bath is weakly dependent on $K$, which is the same as the NRG results. However, the boundary, located at $\alpha_c=1/8+\mathcal{O}(\Delta/\rm\omega_c)$, is also weakly dependent on $\Delta$, a result at variance with the NRG counterpart of $\alpha_c=0.15+\mathcal{O}(\Delta/\rm\omega_c)$. For the sub-Ohmic bath, our calculated $\alpha_c$ is in good agreement with that of the NRG approach for the whole range of $K$ values.

Figure~3(a) shows the difference in the ground
state energy between our calculation and that in \cite{nazir}
in the presence of an Ohmic bath ($s=1$). For
the delocalized phase $\alpha<\alpha_c$, our $E_g$ is the same as
that of Ref.~\cite{nazir}. However, above the transition point
$\alpha\ge\alpha_c$, the lower ground state energy indicates that
the ansatz of this work is a better one for the real ground state.
As shown in the figure the calculated value of the parameter $\sigma_0$,
is nearly zero for the delocalized phase
($\alpha<\alpha_c$), but increases quickly above the transition point.

Figure~3(b) shows the ground state average of $\langle\sigma^x\rangle$
and the renormalized bias $\epsilon'$ as functions of $\alpha$ for
an Ohmic bath. One can see that our calculated average
$\langle\sigma^x\rangle$ (see Eq.(24)) is the same as that of
Ref.~\cite{nazir} for $\alpha< \alpha_c$, and in this regime, the
renormalized bias $\epsilon'\approx\epsilon$ is very small, while for
$\alpha \ge \alpha_c$, $\epsilon'$ increases quickly. Since our
interest is mainly on the QCP, our calculation is restricted to the
parameter regime of $0<\alpha\le 1.1\alpha_c$ where
$\epsilon'/\omega_c<0.05$ and the transition term $|G\rangle\langle
D|+|D\rangle\langle G|$ in Eq.(19) can be safely neglected.

Eqs.~(25) and (26) are used to get the ground state averages of
$\langle\sigma^z\rangle=\langle\sigma^z_1\rangle=\langle\sigma^z_2\rangle$
as a function of $\epsilon$, $\alpha$, $\Delta$, or $K$.
As critical exponents are the most interesting QPT properties,
we first consider a critical exponent $\delta$ defined by
\begin{eqnarray}
&&\langle\sigma^z\rangle\sim \epsilon^{1/\delta},
\end{eqnarray}
where $\alpha$, $\Delta$, and $K$ are kept fixed at their critical
values.  Fig.~4 shows a log-log plot of the relation between
$\langle\sigma^z\rangle$ and $\epsilon/\omega_c$ for
$\Delta/\omega_c=0.1$ and $K=0$, and $\alpha=\alpha_c$. A series of
$s$ values are taken. The filled blue dots in Fig.~2 indicate the
transition points in the phase diagram where we cross the phase
boundary to calculate the curves in Fig.~4. One can see the
power-law scaling over more than two orders of magnitude, and the
critical exponent $\delta$ can be determined from simply fitting the
slope. The fitting results are listed in the second column of Table
1, which are in the close vicinity of $\delta=3$.

Second, the static susceptibility is related to the critical
exponent $\gamma$,
\begin{eqnarray}
&&\chi=\left.\frac{\langle\sigma^z\rangle}{\epsilon}\right
|_{\epsilon\to 0}\sim \frac{1}{(\alpha_c-\alpha)^{\gamma}},
\end{eqnarray}
where $\Delta$ and $K$ are kept fixed. Figure 5 shows a log-log plot
of the relation between $\chi$ and $\alpha_c-\alpha$ for various
values of $s$ (the transition points are again the filled blue dots
in Fig.~2). There is a power-law scaling and the critical exponent
$\gamma$ can be determined from simply fitting the slope. The
fitting results, which are listed in the third column of Table 1,
are found to be quite close to the value of $\gamma=1$.

Another three critical exponents are defined as follows,
\begin{eqnarray}
&&\langle\sigma^z\rangle\sim (\alpha-\alpha_c)^{\beta},\\
&&\langle\sigma^z\rangle\sim (\Delta_c-\Delta)^{\beta'},\\
&&\langle\sigma^z\rangle\sim (K_c-K)^{\zeta} .
\end{eqnarray}
They can be determined in the similar way, that is, by simply
fitting the slope in a log-log plot and the results are listed in
the fourth (transition points are filled blue dots in Fig.~2), fifth
(transition points are filled blue dots in Fig.~1), and sixth
(transition points are  red circles in Fig.~2) columns of Table 1.
All these fitted exponents are found to be close to $1/2$.

We have checked that these extracted exponents are independent of
the position in the phase diagram where the phase boundaries are
crossed. We note that our transformed Hamiltonian
$H'_0+U_{\epsilon}$ is a two-site Ising model in both the transverse
($\eta\Delta$) and longitudinal ($\epsilon'$) field. For the lattice
Ising model (one-, two-, and three-dimensional) in transverse field
it is well-known that there is a quantum phase transition when the
transverse field changes across some critical value\cite{tim}. It
was proved that the critical exponents of the d-dimensional Ising
model in transverse field are the same as those of the classical
Ising model (without the transverse field) in (d+1)-dimension. In
mean-field approximation the critical exponents of the quantum Ising
model (in transverse field) are $\delta=3$, $\gamma=1$, and
$\beta=1/2$, which are independent of the lattice dimension and the
coordination number, and different from the exact analytic solution
(for one-dimension) and numerical exact solutions (Monte Carlo,
renormalization group, etc.). Note that these mean-field critical
exponents are the same as our values for the two-qubit SBM. This is
an indication that our theory for the QPT of the two-qubit SBM is a
mean-field theory, that is, the effect of quantum fluctuations has
been taken into account by a self-consistent mean-field.

Here we explain briefly how our mean-field approximation works. The
two-qubit system and the heat bath are decoupled by the unitary
transformation and in the generator $S$ of the transformation we
introduce two ``mean-field'' displacement of oscillators: (1)the
dynamic displacement $\xi_k(\sigma^z_1+\sigma^z_2)$ related to the
high-frequency oscillators since $\xi_k\approx 1$ for large
$\omega_k$, which modifies the original tunneling $\Delta\to
\eta\Delta$ (Eq.~(8)) and renormalizes the Ising coupling $K\to K-V$
(Eq.~(9)); (2)the static displacement $(1-\xi_k)\sigma_0$ related to
the lower-frequency oscillators as $1-\xi_k\approx 1$ for
$\omega_k\to 0$, which leads to the modified bias
$\epsilon\to\epsilon'$ (Eq.~(5)). As shown above, self-consistent
calculations have been carried out to determine these modified
parameters and to include the effect of the quantum fluctuations.

Moreover, all the critical exponents listed in Table 1 are
independent of the bath index $s$ and this is a feature similar to
the mean-field exponents of the quantum Ising model which are
independent of the dimension and the coordination number. We note
that, for $s=1/2$, our critical exponents are the same as the
scaling analysis result of Ref.\cite{hur}.

As for the critical exponents, our results come from a self-consistent mean-field ground state. It leads reasonably to s-independent plain mean-field critical exponents. In contrast, the critical exponents of the mean-field analysis in Ref. \cite{hur} is based on the quantum to classical mapping of the spin-boson model to the one-dimensional classical Ising model with long-range interaction $J_{ij}=J/|i-j|^{1+s}$, which results in the s-dependent critical exponents. On the other hand, as pointed out in Ref.~[1], the NRG is not well suited to describe the system close to the transition for $s < 1/2$, and the calculation is therefore restricted to $s \geq 1/2$. It is our belief that it is not accidental that the critical exponents $\zeta(s=1/2)=1/2$ and $\beta(s=1/2)=0.5$ of the NRG are equivalent to those of our theory. Recent Quantum Monte Carlo simulation yields classical exponents of $\gamma=1$ and $\beta=0.5$ for $s < 1/2 $ in a multi-qubit SBM\cite{winter}, but for $ s> 1/2$, their critical exponents are dependent on $s$ while ours are independent of $s$.

\section{The entanglement entropy}

The reduced system density matrix $\rho_S$ is given by tracing the
total (system + bath) density operator over the boson bath:
$\rho_S=\mbox{Tr}_B[\rho_{SB}]$. If the ground-state reduced density
matrix of the two-qubit system $\rho_S$ is known, the von Neumann
entanglement entropy can be calculated from $\rho_S$: ${\cal
E}=-\mbox{Tr}[\rho_S\rm\log_{2}\rho_S]$ \cite{zz,hur}. From the trial
ground state (24) we have
\begin{eqnarray}
&&\rho_{SB}=|g.s.\rangle \langle g.s.|=\exp(-S)|G\rangle
|\{0_k\}\rangle\langle\{0_k\}|\langle G|\exp(S).
\end{eqnarray}
Thus,
\begin{eqnarray}
&&\rho_{S}=\mbox{Tr}_B\{\exp(-S)|G\rangle
|\{0_k\}\rangle\langle\{0_k\}|\langle G|\exp(S)\}.
\end{eqnarray}
Note that there are both the spin operators $\sigma^z$ and the
bosonic operators $b^{\dag}_k-b_k$ in $S$. For the trace operation
over the bath ($\mbox{Tr}_B$) we use Eqs. (17), (18), (10), (11) and
$|1\rangle=(|\uparrow \rangle+|\downarrow\rangle)/\sqrt{2}$,
$|2\rangle=(|\uparrow \rangle-|\downarrow\rangle)/\sqrt{2}$
($|\uparrow (\downarrow)\rangle$ is the eigenstate of $\sigma^z$:
$\sigma^z|\uparrow
(\downarrow)\rangle=+(-)|\uparrow(\downarrow)\rangle$) to express
$|G\rangle$ as
\begin{eqnarray}
&&|G\rangle =\cos\theta |A\rangle+\sin\theta|B\rangle\nonumber\\
&&=\frac{1}{\sqrt{2}}\{(u\cos\theta+\sin\theta)|\uparrow\uparrow\rangle+
(u\cos\theta-\sin\theta)|\downarrow\downarrow\rangle
+v\cos\theta[|\uparrow\downarrow\rangle+|\downarrow\uparrow\rangle]\}.
\end{eqnarray}
Then,
\begin{eqnarray} \label{rou}
&&\exp(-S)|G\rangle
=\frac{1}{\sqrt{2}}(u\cos\theta+\sin\theta)\exp(-S_+)|\uparrow\uparrow\rangle\nonumber\\
&&+
\frac{1}{\sqrt{2}}(u\cos\theta-\sin\theta)\exp(-S_-)|\downarrow\downarrow\rangle
+v\cos\theta\exp(-S_0)\frac{1}{\sqrt{2}}[|\uparrow\downarrow\rangle+|\downarrow\uparrow\rangle]\},
\end{eqnarray}
where
\begin{eqnarray}
&&S_+=\sum_k(f_k+\frac{g_k\xi_k}{\omega_k})(b^{\dag}_k-b_k),~S_-=\sum_k(f_k-\frac{g_k\xi_k}{\omega_k})(b^{\dag}_k-b_k),~
S_0=\sum_kf_k{\omega_k}(b^{\dag}_k-b_k),\nonumber\\
\end{eqnarray}
and $f_k=g_k(1-\xi_k)\sigma_0/2\omega_k$. Now there are no system
operators in $S_+$, $S_-$ and $S_0$ so that the cyclic properties of
the trace can be used for trace operation in Eq.~(36),
\begin{eqnarray}
&&\rho_S= \\
&&\left(
\begin{array}{cccc}
{1\over 2}(u\cos\theta+\sin\theta)^2 & \frac{v\eta}{\sqrt{2}}\cos\theta(u\cos\theta+\sin\theta) & {1\over 2}(u^2\cos^2\theta-\sin^2\theta)\eta^4 & 0\\
\\
\frac{v\eta}{\sqrt{2}}\cos\theta(u\cos\theta+\sin\theta) & v^2\cos^2\theta & \frac{v\eta}{\sqrt{2}}\cos\theta(u\cos\theta-\sin\theta) & 0\\
\\
{1\over 2}(u^2\cos^2\theta-\sin^2\theta)\eta^4 &
\frac{v\eta}{\sqrt{2}}\cos\theta(u\cos\theta-\sin\theta) & {1\over
2}(u\cos\theta-\sin\theta)^2 & 0\\
\\ 0 & 0 & 0 & 0
\end{array}
 \right ).\nonumber
 \end{eqnarray}
Because of the decoupling of the ``dark" state
$\frac{1}{\sqrt{2}}[|\uparrow\downarrow\rangle-|\downarrow\uparrow\rangle]$,
all elements of the density operator $\rho_S$ in the bottom row and
right column are $0$. If we know the three eigenvalues of the upper
left $3\times 3$ sub-matrix then the entanglement entropy is,
\begin{eqnarray}
&&{\cal E}=-\sum_{i=1}^{3} \lambda_i\log_2\lambda_i,
\end{eqnarray}
where $\lambda_i$ ($i=1,2,3$) are the eigenvalues of the $3\times 3$
sub-matrix. As the trace of the density operator is
$\mbox{Tr}_S\rho_S=1$, it is easy to prove that $0\le{\cal E}\le
2$\cite{hur}. ${\cal E}=0$ indicates the absence of entanglement
between the qubits and the bath.

The eigenvalues of $\rho_S$ can be calculated numerically. The
entanglement entropy $\cal E$ for the Ohmic case of $s=1$ is shown
in Fig.~6(a) as a function of the coupling strength $\alpha$ for
three values of tunneling $\Delta$ (we set $K=0$ and
$\epsilon/\omega_c=10^{-6}$). When $\alpha=0$ there is no
entanglement between qubits and environment and ${\cal E}=0$. The
entanglement entropy increases with increasing $\alpha$ in the
delocalized phase, reaches a plateau and then drops quickly to zero
at the transition point $\alpha=\alpha_c$ (Here and in the following
figures our calculation is restricted to the range $0<\alpha\le
1.1\alpha_c$ because in this range $\epsilon'/\omega_c<0.05$ and the
transition term $|G\rangle\langle D|+|D\rangle\langle G|$ in Eq.(19)
can be safely dropped). As pointed out in Ref.~\cite{hur}, the
plateau indicates that coherence is lost prior to localization, that
is, it shows that the system is in the coherent to incoherent
crossover before final trapping in the localized phase.

Figure~6(b) displays the entanglement entropy $\cal E$ for the sub-Ohmic
case of $s=1/2$ and three values of tunneling $\Delta$ (we set $K=0$ and
$\epsilon/\omega_c=10^{-6}$). Obviously, for the sub-Ohmic bath the
entanglement entropy reaches a sharp peak right at the
transition point and there is no plateau corresponding to the
coherent to incoherent crossover.

Figure~6(a) is corresponding to the case of $K=0$, then the renormalized Ising
coupling is $-V\sigma^z_1\sigma^z_2$. In Fig.~7 we check the
$\cal E$ versus $\alpha$ relation for finite values of the Ising coupling $K$
($s=1$, $\Delta=0.1$, $\epsilon/\omega_c=10^{-6}$). From Fig. 7(a), we observe that
as $K$ changes from the ferromagnetic ($K<0$) to the
antiferromagnetic ($K>0$, note that the renormalized Ising coupling
is $K-V$) the width of the plateau is reduced considerably, and a spike emerges instead for large positive
values for $K\ge 0.25\omega_c$. This indicates that the localization
transition occurs right next to the regime where spin
dynamics is coherent \cite{hur}, and coherence is lost in a manner similar to the sub-Ohmic case of Fig.~6(b).
In Figs.~7(b) and 7(c), we show the comparison of the NRG results with ours. For several values of $K$, the slopes of our scaled data are similar to those of the NRG approach.

Figure.~8 shows the entanglement entropy $\cal E$ as a function of
$\alpha$ for various values of Ising coupling $K$ in the sub-Ohmic regime of $s=1/2$
(we set $\Delta=0.1$ and $\epsilon/\omega_c=10^{-6}$). There is a sharp peak
at the transition point for both the ferromagnetic ($K<0$) and the
antiferromagnetic ($K>0$) Ising coupling, but the width of the peak
of the ferromagnetic coupling is much smaller than that of the
antiferromagnetic one.
In Figs. 8(b), 8(c), and 8(d), we show the comparison of the NRG results with ours. For several values of $K$, it is found that the slopes of our scaled data agree well with those of the NRG approach.

In Fig.~9 we show the ${\cal E}$ versus $\alpha$ relations for $s=1/4$, $1/2$, $3/4$, $9/10$, and $1$ (from left to right). Here we set Ising coupling $K=0$. One can see that with increasing index $s$ a sharp peak ($s=1/4$) at the transition point changes gradually (with $s=1/2$, $3/4$, $9/10$) to a plateau ($s=1$) on the left side of the peak.

\section{Qubit-qubit correlation}
In this section, in order to investigate the correlation
between the two qubits mediated by the common bath and the effects of direct Ising couple,
we calculate the qubit-qubit correlation function of the ground state. It is defined as
\begin{eqnarray}
&&{C_{12}}= \langle \sigma^z_{1}\sigma^z_{2}\rangle-\langle \sigma^z_{1}\rangle\langle \sigma^z_{2}\rangle,
\end{eqnarray}
where $\langle \bullet \rangle=\langle g.s.|\bullet  |g.s.\rangle$. By the reduced density matrix Eq. (\ref{rou}), we immediately arrive at
\begin{eqnarray}
&&{C_{12}}= (u^2-v^2)\cos^2\theta+\sin^2\theta -\frac{1}{4}\sigma_0^2.
\end{eqnarray}
In Fig. 10 we show the correlation function $C_{12}$ for different bath indexes $s$. In Figs. 10(a) and 10(b), we show our calculated results and the data of the QMC simulations for $K=0$\cite{winter}. Due to the coupling of the qubits and bath, there is an indirect Ising coupling $-V$. The function $\langle \sigma^z_{1}\sigma^z_{2}\rangle$ is nonzero even in the delocalized phase due to the effective
ferromagnetic interaction mediated by the common bath. It is obvious to see that the the fluctuation increases with the increase of the dissipative coupling before the QPT. At the critical point $\alpha_c$, $C_{12}$ is reached the maximum value, which means that the QPT happens. After passing $\alpha_c$, $C_{12}$ decreases rapidly. By comparison, our results are in good agreement with the QMC results, especially for the deep sub-Ohmic bath $s \leq 1/2$. In Fig. 10(a), for the Ohmic bath, our results of the delocalized phase agree well with the QMC data\cite{winter}.  In Fig. 10(b), we notice that the transition of our results occurs at $\alpha_c=0.133$ while that of the QMC happens at $\alpha \approx 0.175$.

In Figs. 10(c) and 10(d), we show the effects of direct Ising couple $K$ on the correlation function for $s=1$ and $s=1/2$, respectively. For the Ohmic case the $C_{12}$ has a character of plateau at the lower dissipation in the ferromagnetic case $K<0$, while for larger values of $K$ the plateau shrinks to a peaklike structure.  It is clearly seen that the peak value of $C_{12}$ for the antiferromagnetic situation is much higher than those for the ferromagnetic case. For the sub-Ohmic case $s=1/2$, the $C_{12}$ exhibits a character of cusp for any $K$, similar to the entanglement entropy in Fig. 8(a).

\section{Discussion and conclusion}

We have proposed an ansatz to study a two-qubit system interacting
with a dissipative environment in the ground state, and it is shown
that, as a result of the competition between the intrasite tunneling
and the intersite correlation, a quantum phase transition to the
localized phase may occur at some critical coupling constant
$\alpha_c$. By calculating the ground state entanglement entropy
between the qubits and the bath as well as the qubit-qubit correlation function, we have explored the nature of the
QPT between the delocalized and localized state.

The same criterion as that used in Ref.\cite{hur} is used to
determine the critical coupling in this work, that is, the emergence
of a non-zero ground state expectation of $\langle\sigma^z\rangle$
as the coupling $\alpha$ increases across some critical point
$\alpha_c$. For the two-qubit system in Ohmic bath we get
$\alpha_c=1/8+O(\Delta/\omega_c)$ which is quite close to the NRG
result $\alpha_c=0.15+O(\Delta/\omega_c)$\cite{hur}. But the
criterion used in Ref.\cite{nazir} is the vanishing of the
renormalized tunneling $\eta\to 0$, which leads to $\alpha_c=0.5$
for the two-qubit system in Ohmic bath. However, for the
single-qubit system in Ohmic bath both the criteria $\eta\to 0$ and
$\langle\sigma^z\rangle\neq 0$ give the same critical value
$\alpha_c=1$, at least in the scaling limit $\Delta/\omega_c\to
0$(Refs.[2-12]). This difference comes from the two-qubit
correlation and the renormalized Ising coupling $V$ (Eq.(9)) which
shift the QCP of the two-qubit SBM substantially as compared to that
of the single-qubit case.

A new unitary transformation has been utilized, in which a
$\omega_k$-dependent function $\xi_k$ is introduced and the
functional form of it is determined by setting zero the matrix
element of $H'_1$ between the ground state and the lowest-lying
excited state of $H'_0+U_{\epsilon}$. Then we get the ground state
$|G\rangle |\{0_k\}\rangle$ for the transformed Hamiltonian
$H'_0+U_{\epsilon}+H'_1$ ($H'_1|G\rangle |\{0_k\}\rangle=0$) with
the ground state energy Eq.~(22). Generally speaking, our approach
is to decouple the two-qubit system from the heat bath by the
unitary transformation with the generator $S$ (Eq.(2)). In $S$ we
introduce two ``mean-field'' displacement of oscillators: (1)The
dynamic displacement $\xi_k(\sigma^z_1+\sigma^z_2)$ related to the
high-frequency oscillators since $\xi_k\approx 1$ for large
$\omega_k$, which modifies the original tunneling $\Delta\to
\eta\Delta$ (Eq.~(8)) and renormalizes the Ising coupling $K\to K-V$
(Eq.~(9)). (2)The static displacement $(1-\xi_k)\sigma_0$ related to
the lower-frequency oscillators as $1-\xi_k\approx 1$ for
$\omega_k\to 0$, which leads to the modified bias
$\epsilon\to\epsilon'$ (Eq.~(5)). Self-consistent mean-field
calculations have been carried out to determine these modified
parameters, then the effect of the quantum fluctuations is included.
Our calculated critical exponents are the same as the mean-field
critical exponents of the Ising model in transverse field.

In our work, the unperturbed part of the transformed Hamiltonian
$H'_0+U_{\epsilon}$ can be solved exactly, but nonetheless contains
the main physics of the two-qubit SBM. For the ground state the
first-order Hamiltonian $H'_1$ can be neglected because
$H'_1|G\rangle |\{0_k\}\rangle=0$. The main approximation in our
treatment is the omission of $H'_2$ (Eq.~(7)). The reason to justify
this approximation is that, since $\langle\{0_k\}|\langle
G|H'_{2}|G\rangle|\{0_k\}\rangle=0$ (because of the definition for
$\eta$ (Eq.(8)), the terms in $H'_{2}$ are related to the
multi-boson non-diagonal transitions (like $b_k b_{k'}$ and
$b^{\dag}_k b^{\dag}_{k'}$). The contributions of these non-diagonal
terms to the ground state energy are $O(g^2_kg^2_{k'})$ and higher.
For the ground state the contribution from these multi-boson
non-diagonal transition may be dropped safely. We have made
substantial arguments in our previous publication\cite{zheng2} that
this omission is justified.

\vskip 0.5cm

{\noindent {\large {\bf Acknowledgement}}}

This work was supported by the National Natural Science Foundation
of China (Grants No. 11174198 and 11374208), the National Basic
Research Program of China (Grant No. 2011CB922200), and by the
Singapore National Research Foundation under Project No.
NRF-CRP5-2009-04.

\newpage

\begin{center}
{\Large \bf Figure Captions }
\end{center}

\vskip 0.5cm

\baselineskip 20pt

{\bf Fig. 1}~~~$\alpha$ versus $\Delta$ phase diagram for various
bath type with $K=0$ and very weak bias $\epsilon/\omega_c=10^{-5}$.
The five curves from the top down are for $s=1$, $0.9$, $0.75$,
$0.5$, and $0.25$, respectively. The blue dots indicate the
positions where we cross the phase boundary for calculating the
critical exponents $\beta'$ in the fifth column of Table 1.

\vskip 0.5cm

{\bf Fig. 2}~~~(a)$\alpha$ versus $K$ phase diagram for various bath
type with $\Delta/\omega_c=0.1$ and very weak bias
$\epsilon/\omega_c=10^{-5}$. The five curves from the top down are
for $s=1$, $0.9$, $0.75$, $0.5$, and $0.25$, respectively. The blue
dots indicate the positions where we cross the phase boundary for
calculating the critical exponents in the second ($\delta$), third
($\gamma$), and fourth ($\beta$) columns of Table 1. The red circles
indicate the positions where we cross the phase boundary for
calculating the critical exponents in the sixth column ($\zeta$) of
Table 1. The comparisons of our result and NRG one for $s=1$ and $s=1/2$ are shown in Figs. 2(b) and (c), respectively. Different red symbols stand for the NRG data in Ref.~[1]. The black curves correspond to our calculated data by our ansatz. The short-dash dotted lines in Figs.~ (b) and (c) indicate $K_r=0$.

\vskip 0.5cm

{\bf Fig. 3(a)}~~~The solid line is the difference between our
calculation of the ground state energy and that of \cite{nazir} in
the Ohmic bath $s=1$ with $K=0$, $\Delta/\omega_c=0.1$ and
$\epsilon/\omega_c=10^{-5}$. The dashed-dotted line is the
calculated value of the parameter $\sigma_0$. The arrow at the right
corner is to indicate the transition point $\alpha_c\approx 0.1338$.

\vskip 0.5cm

{\bf Fig. 3(b)}~~~The ground state average of
$\langle\sigma^x\rangle$ and the renormalized bias $\epsilon'$ as
functions of $\alpha$ for Ohmic bath s = 1 with $K=0$,
$\Delta/\omega_c=0.1$ and $\epsilon/\omega_c=10^{-5}$. The solid
line is our result for $\langle\sigma^x\rangle$ and the dashed line
is that of Ref.\cite{nazir}. The arrow at the right corner is to
indicate the transition point $\alpha_c\approx 0.1338$.

\vskip 0.5cm

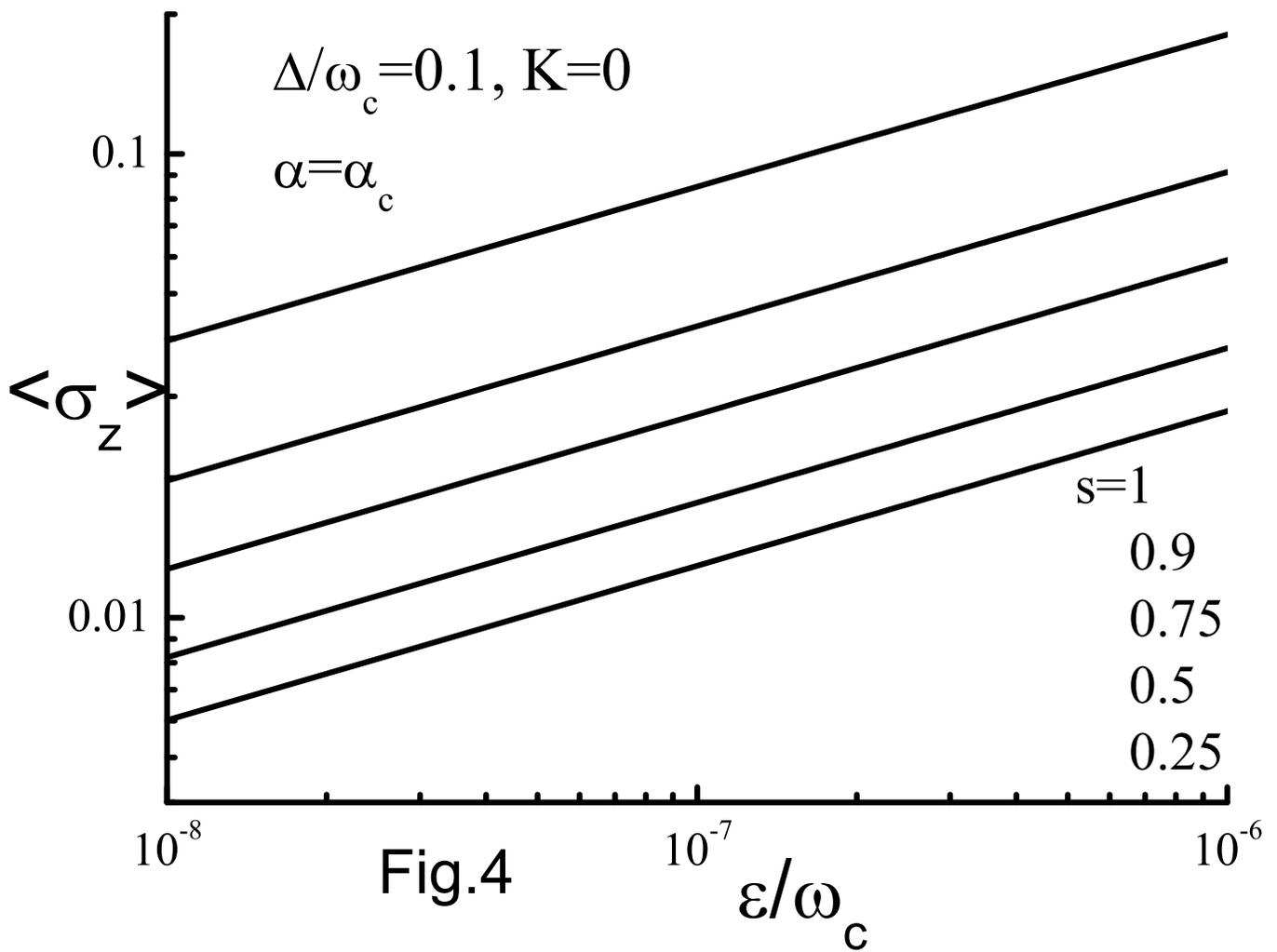
{\bf Fig. 4}~~~The log-log plot of the relation between
$\langle\sigma^z\rangle$ and $\epsilon/\omega_c$ for various bath
type with fixed $\Delta/\omega_c=0.1$, $K=0$ at their corresponding
critical values of $\alpha=\alpha_c$. $s=1$, $0.9$, $0.75$, $0.5$,
and $0.25$ (from top to bottom).

\vskip 0.5cm

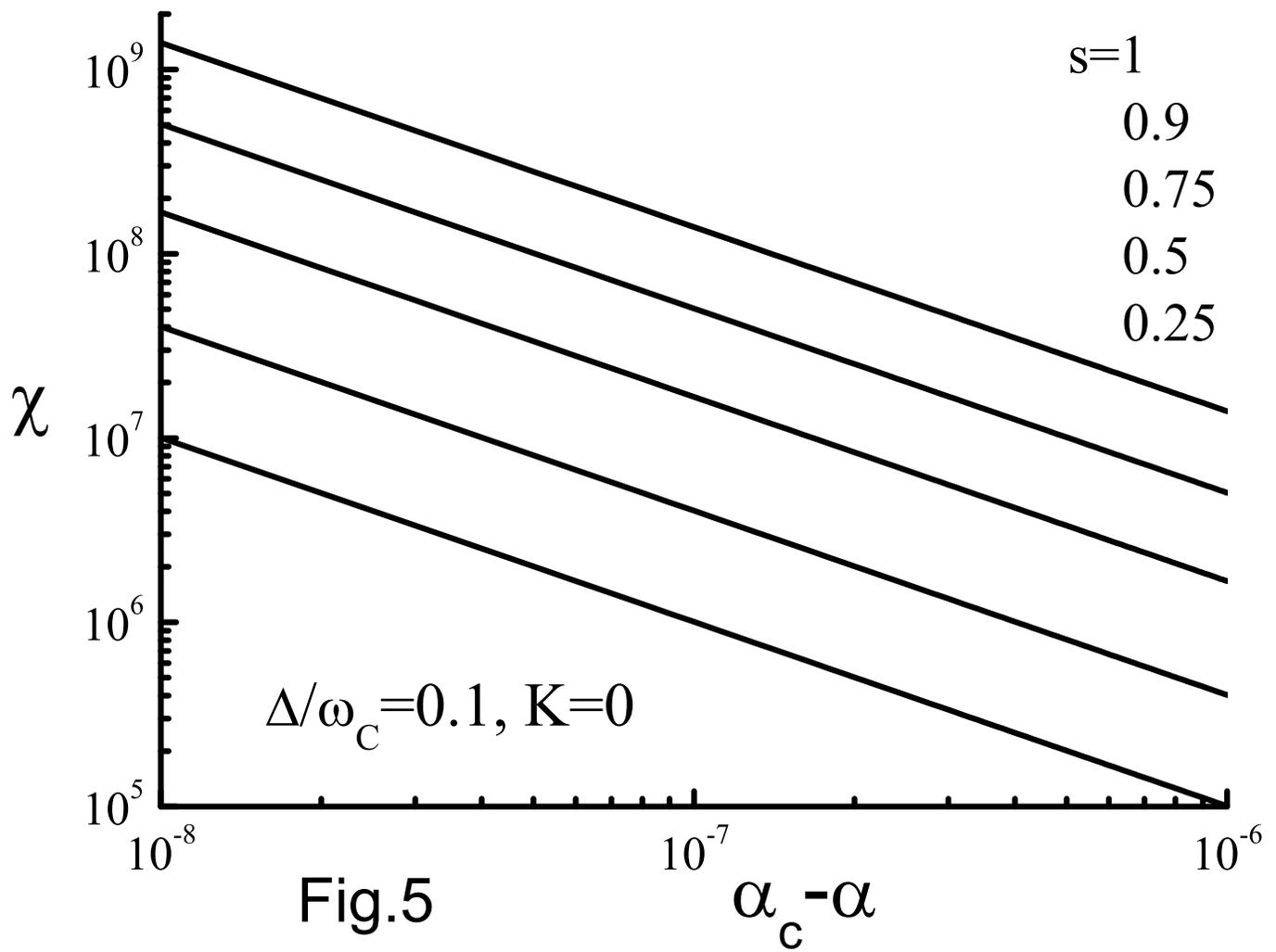
{\bf Fig. 5}~~~The log-log plot of the relation between $\chi$ and
$1/(\alpha_c-\alpha)^{\gamma}$ for various bath type with fixed
$\Delta/\omega_c=0.1$ and $K=0$. $s=1$, $0.9$, $0.75$, $0.5$, and
$0.25$ (from top to bottom).

\vskip 0.5cm

{\bf Fig. 6(a)}~~~The entanglement entropy $\cal E$ as a function of
dissipation $\alpha$ for the Ohmic case of $s=1$ with different
tunneling $\Delta$ ($K=0$, $\epsilon/\omega_c=10^{-6}$).

\vskip 0.5cm

{\bf Fig. 6(b)}~~~The entanglement entropy $\cal E$ as a function of
dissipation $\alpha$ for the sub-Ohmic case of $s=1/2$ with
different tunneling $\Delta$ ($K=0$, $\epsilon/\omega_c=10^{-5}$).

\vskip 0.5cm

{\bf Fig. 7}~~~(a) The entanglement entropy $\cal E$ as a function of
$\alpha$ for different Ising coupling $K$ in Ohmic bath $s=1$
($\Delta=0.1$, $\epsilon/\omega_c=10^{-5}$). The comparisons of the scaled entanglement entropy versus $(\alpha-\alpha_c)/\alpha_c$ for $K=0$ and $4K=0.5\omega_c$ are shown in (b) and (c), respectively.

\vskip 0.5cm

{\bf Fig. 8}~~~(a) The entanglement entropy $\cal E$ as a function of
$\alpha$ for deferent Ising coupling $K$ in sub-Ohmic bath $s=1/2$
($\Delta=0.1$, $\epsilon/\omega_c=10^{-5}$). The comparisons of the scaled entanglement entropy versus $(\alpha-\alpha_c)/\alpha_c$ for $4K/\omega_c=-0.5$, 0 and $0.5$ are shown in (b), (c), and (d), respectively.

\vskip 0.5cm

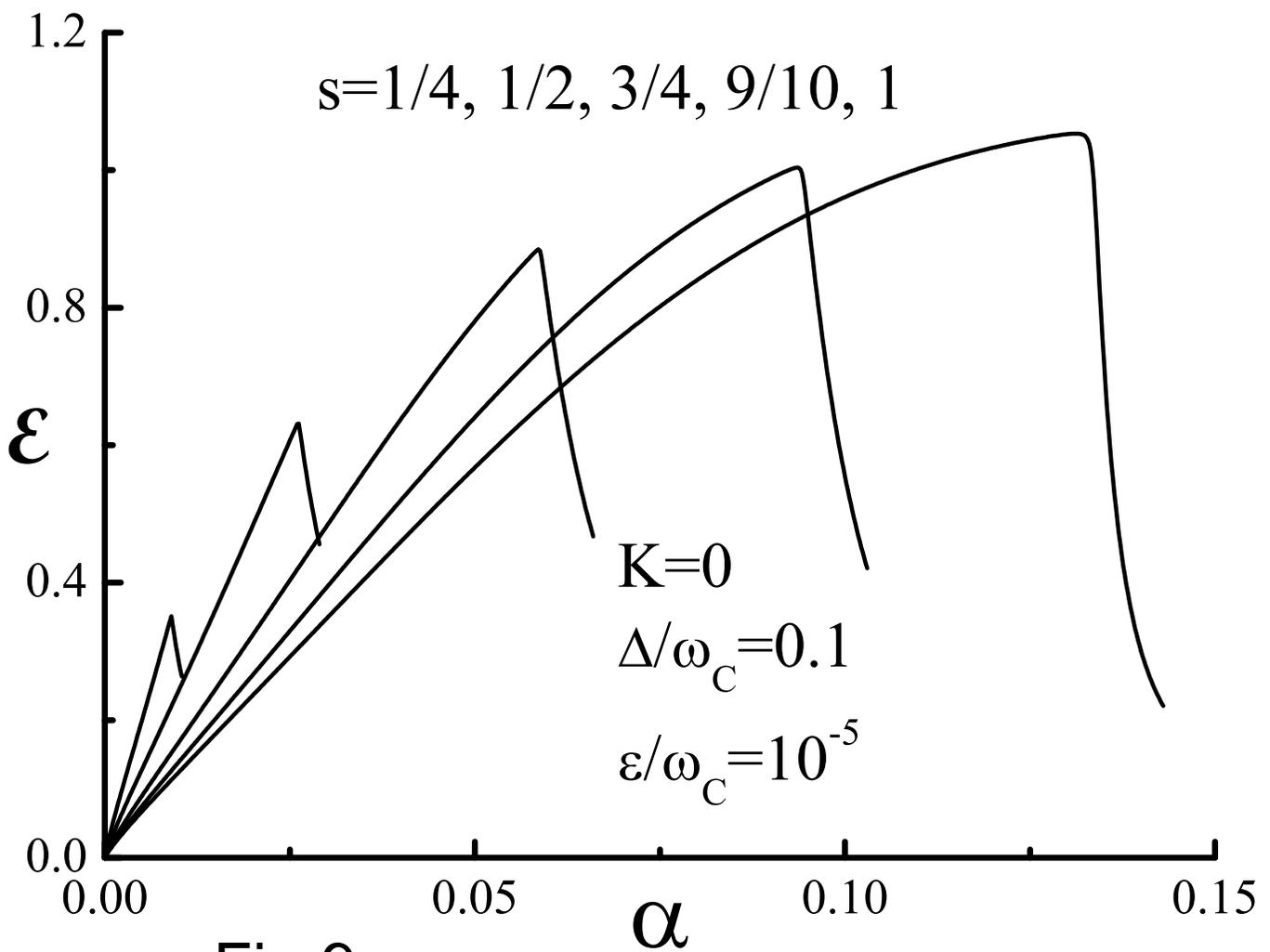
{\bf Fig. 9}~~~The entanglement entropy ${\cal E}$ as a function of
$\alpha$ for the different bath index $s=1/4$, $1/2$, $3/4$, $9/10$,
and $1$ (from left to right).

\vskip 1cm

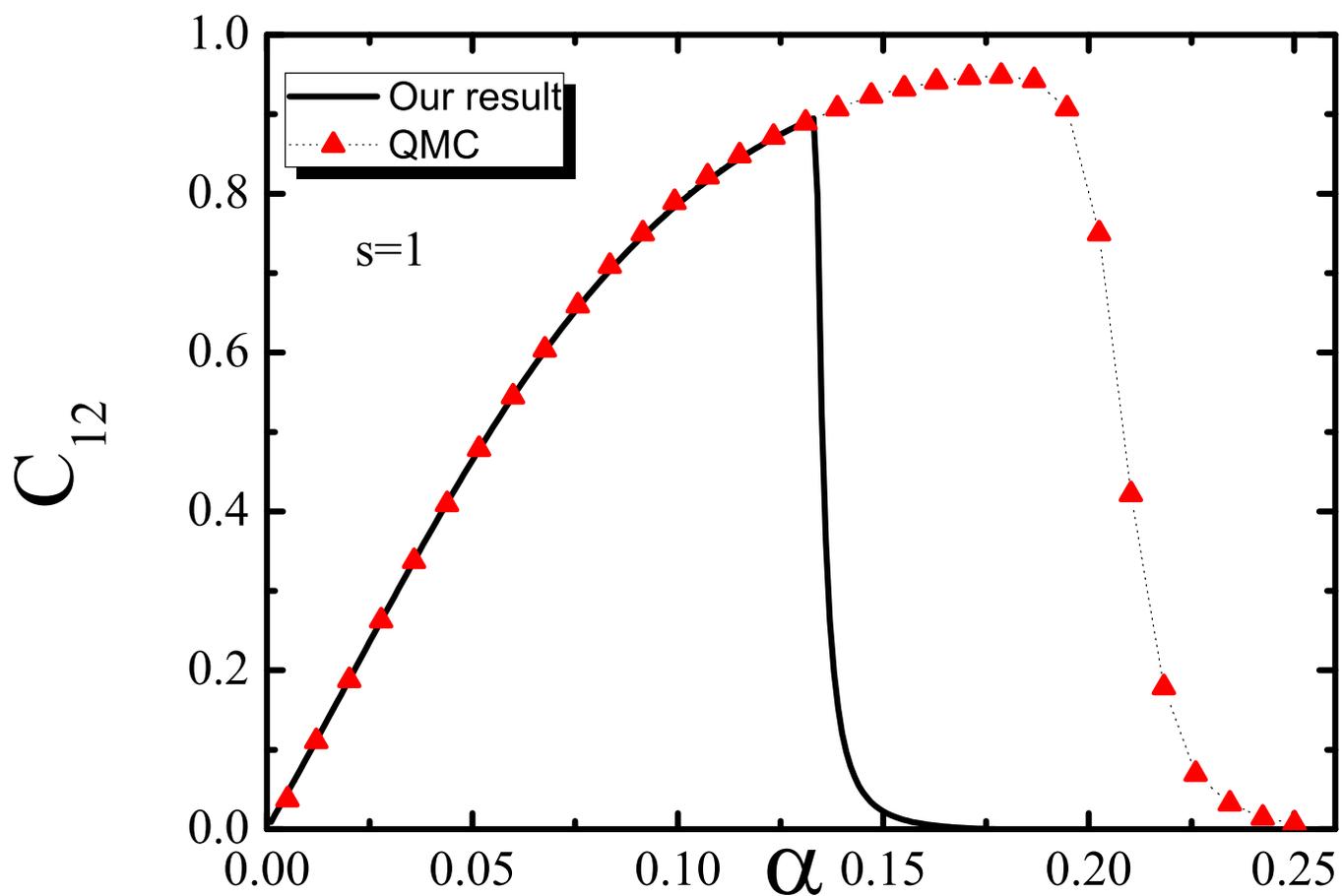
{\bf Fig. 10}~~~(a) The qubit-qubit correlation function $C_{12}$ as a function of
$\alpha$ for Ising coupling $K=0$ with different bath indexes $s=0.25, 0.5, 0.75$, and $1$
($\Delta=0.1$, $\epsilon/\omega_c=10^{-5}$). The data of Quantum Monte Carlo in Ref.[14] are shown for comparison.
(b) $C_{12}$ versus $\alpha$ for $K=0$ in the Ohmic case.
The correlation functions $C_{12}$ for $s=1$ and $s=1/2$ are shown in (c) and (d), respectively.
Different curves are for different values of the Ising coupling $K$.

\vskip 1cm

\begin{center}
{\Large \bf Tables }
\end{center}

{\bf Table 1}~~~Critical exponents of different bath type $s$.

\begin{table}[h]
\centering\setlength\tabcolsep{10pt}
\begin{tabular}{|cccccc|}
\hline
 $s$ & $\delta$  &  $\gamma$ & $\beta$ &
$\beta'$ & $\zeta$
\\ \hline
 0.25    & 3.0009  &  0.99999  &  0.49695  &  0.49988 &  0.49981 \\
 0.5    &  3.0015  &  1.00009  &  0.49848  &  0.49981 &  0.49979 \\
 0.75    & 3.0036  &  1.00041  &  0.49882  &  0.49971 &  0.49980 \\
 0.9   &   3.0088  &  1.00004  &  0.49864  &  0.49960 &  0.49979 \\
 1      &  3.0396  &  0.99999  &  0.49538  &  0.49912 &  0.49965 \\

\hline

\end{tabular}

\end{table}

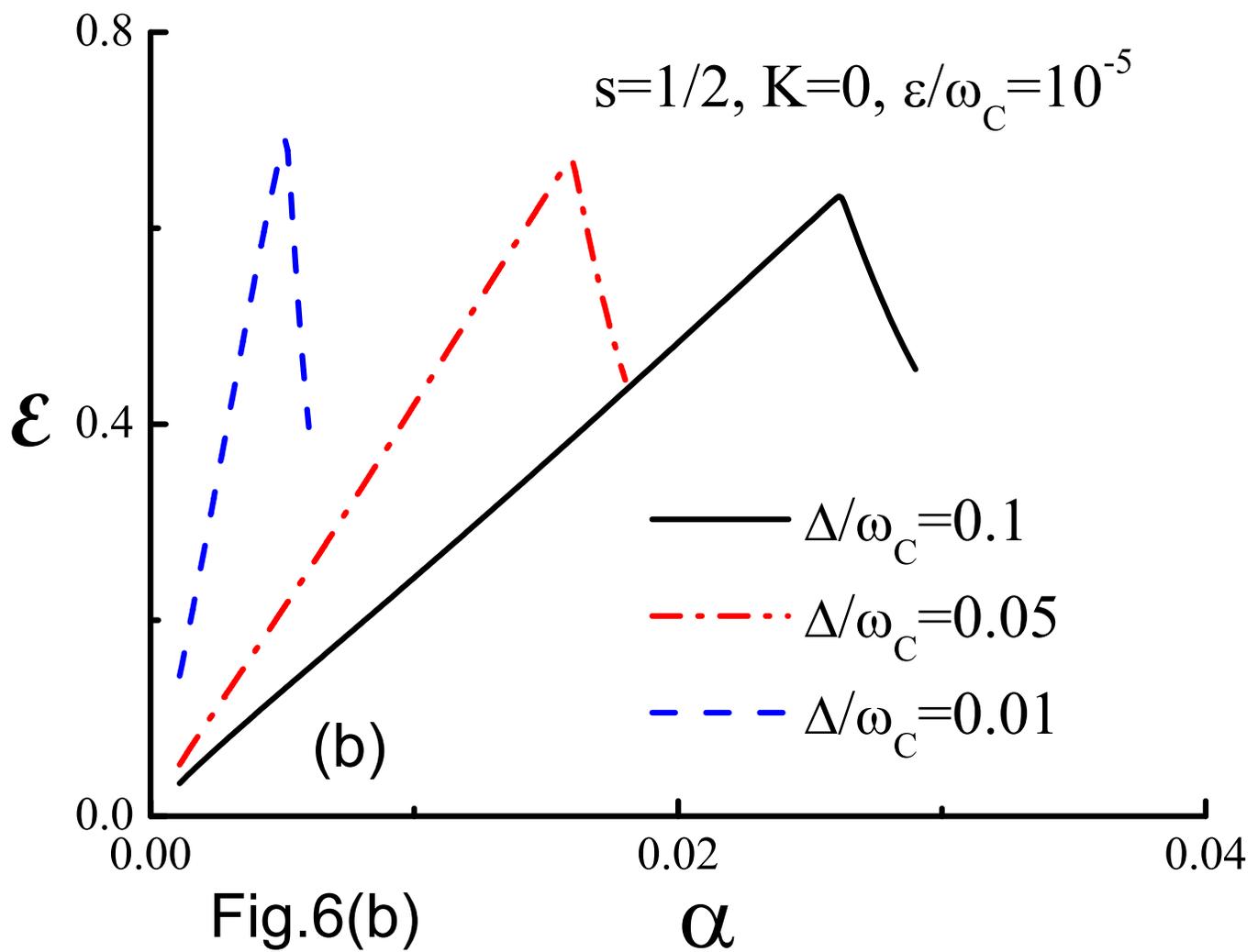
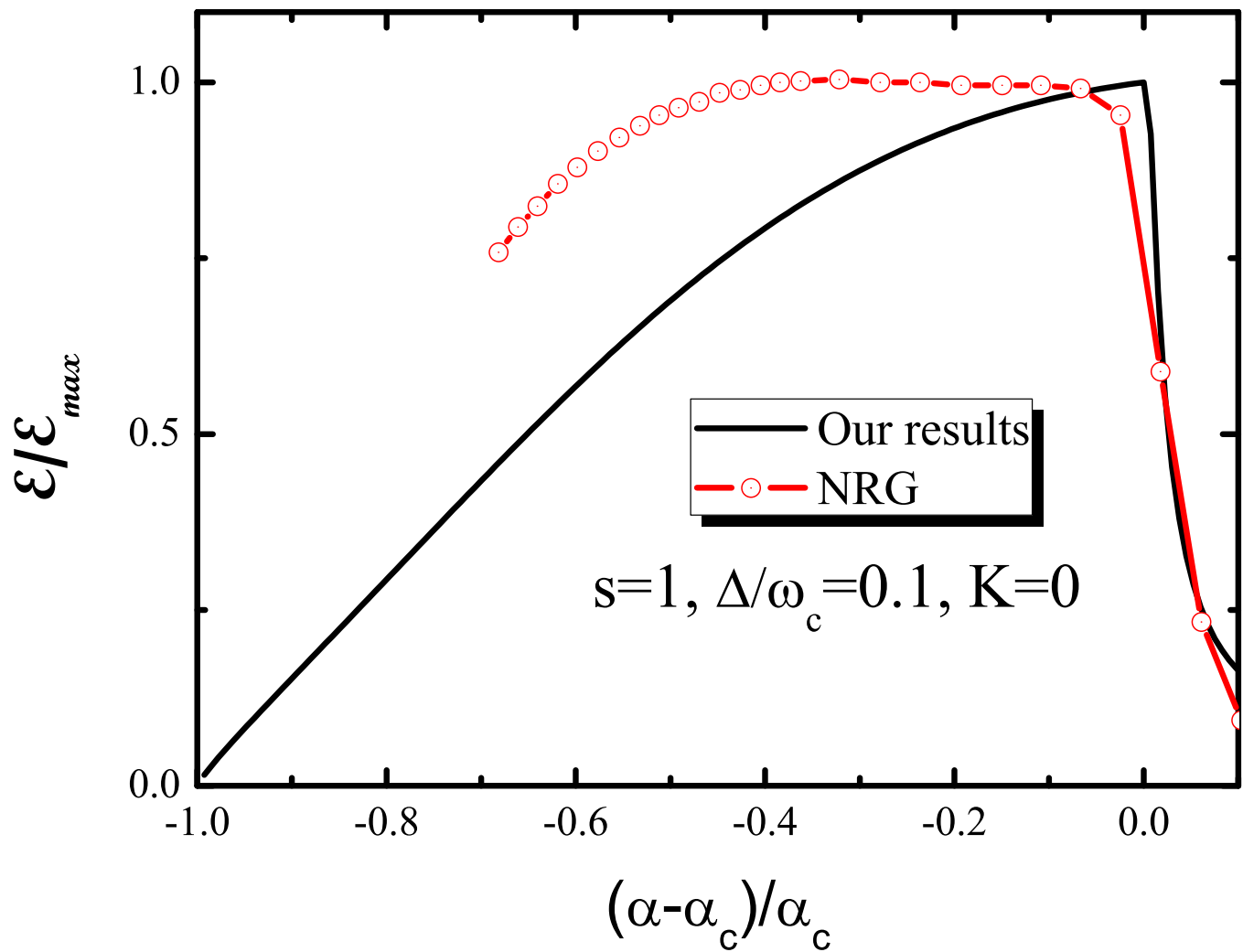
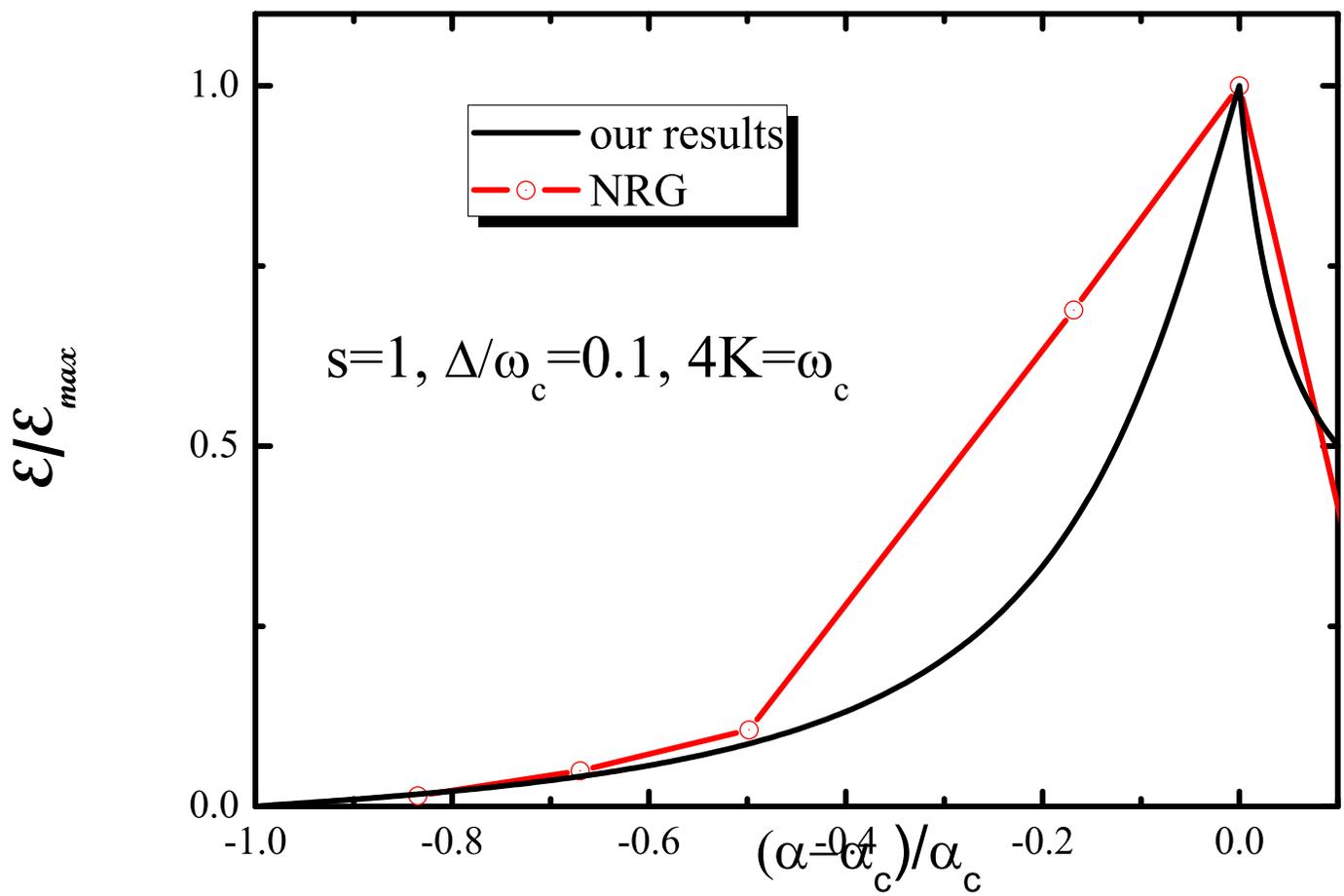
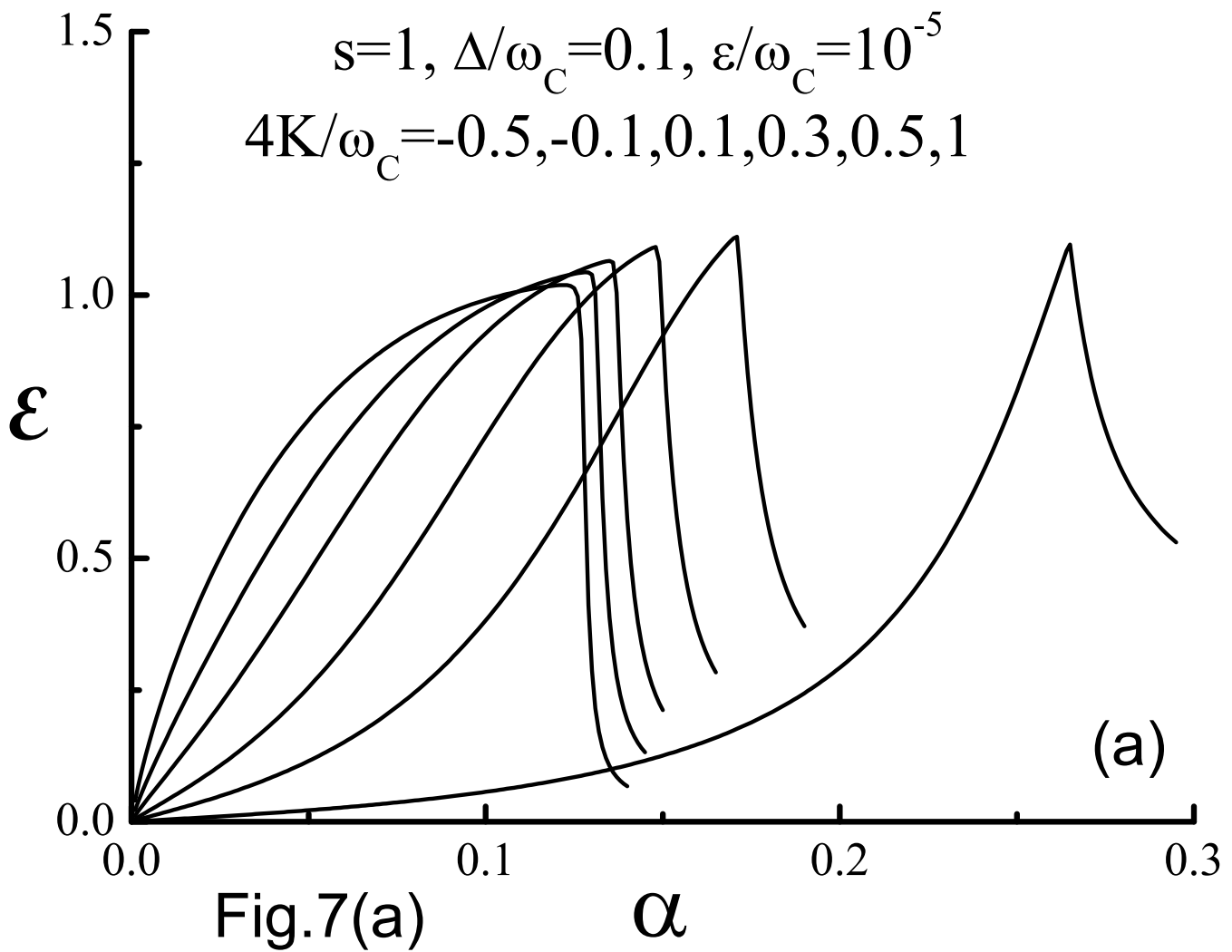
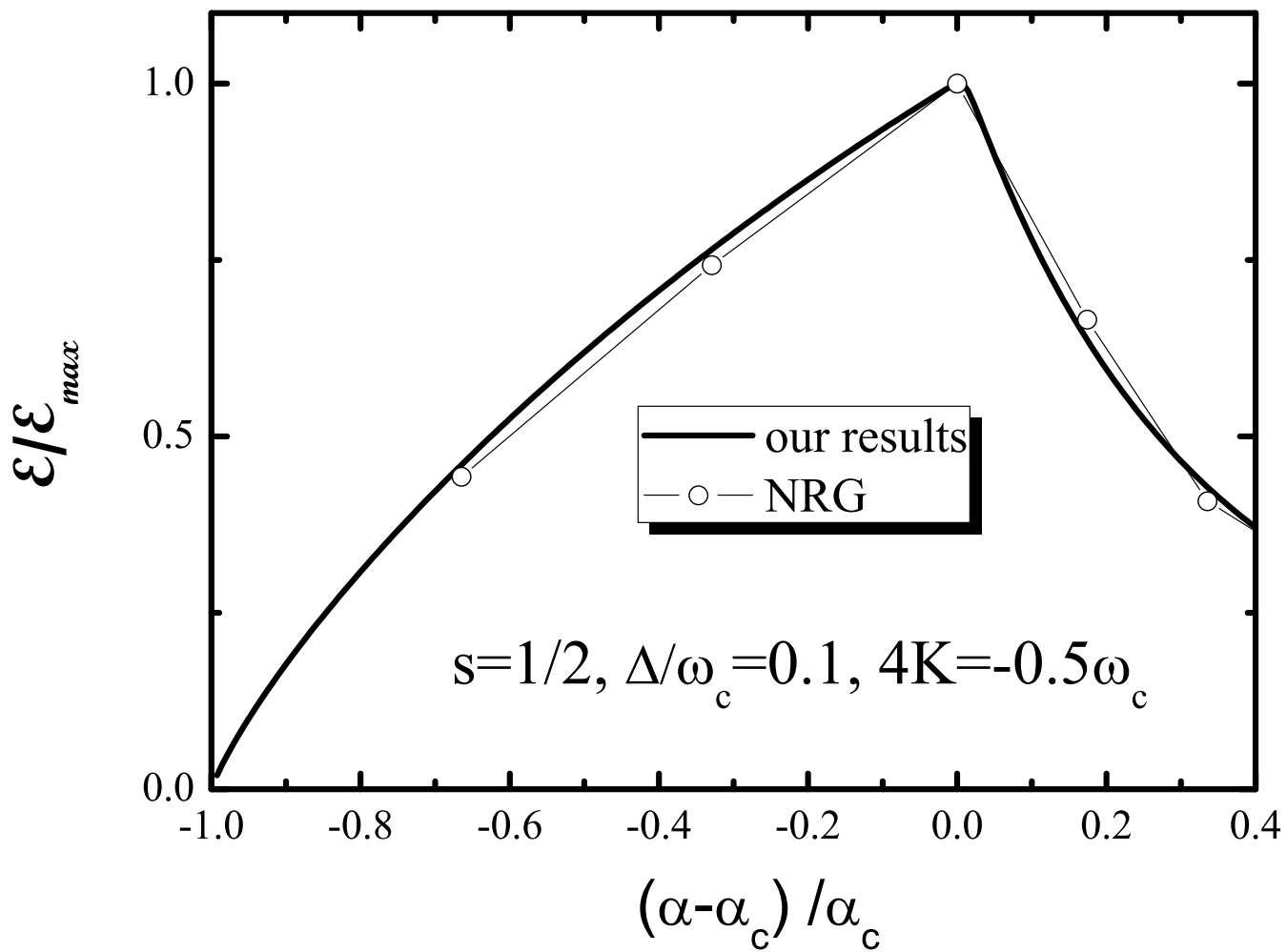
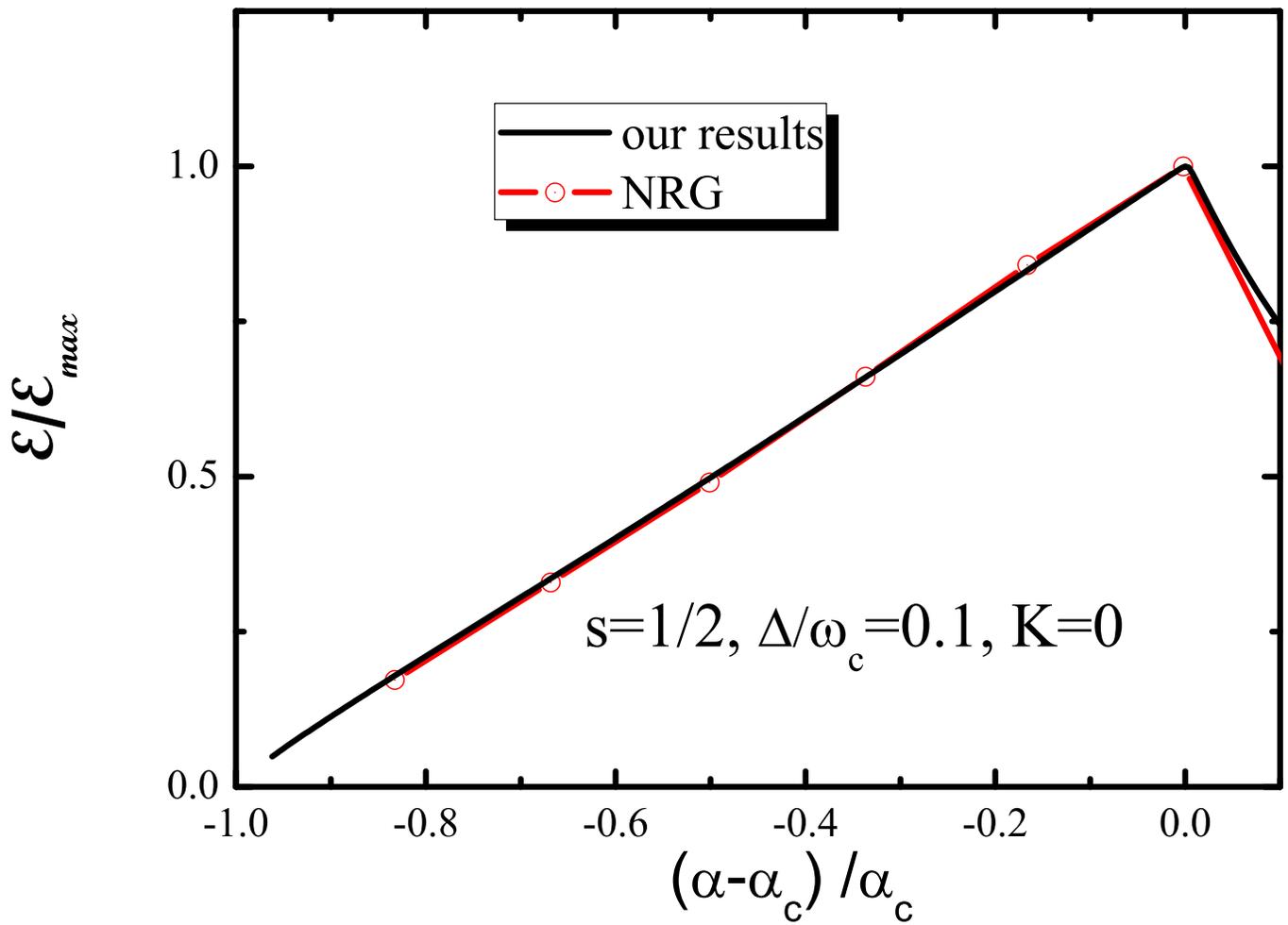
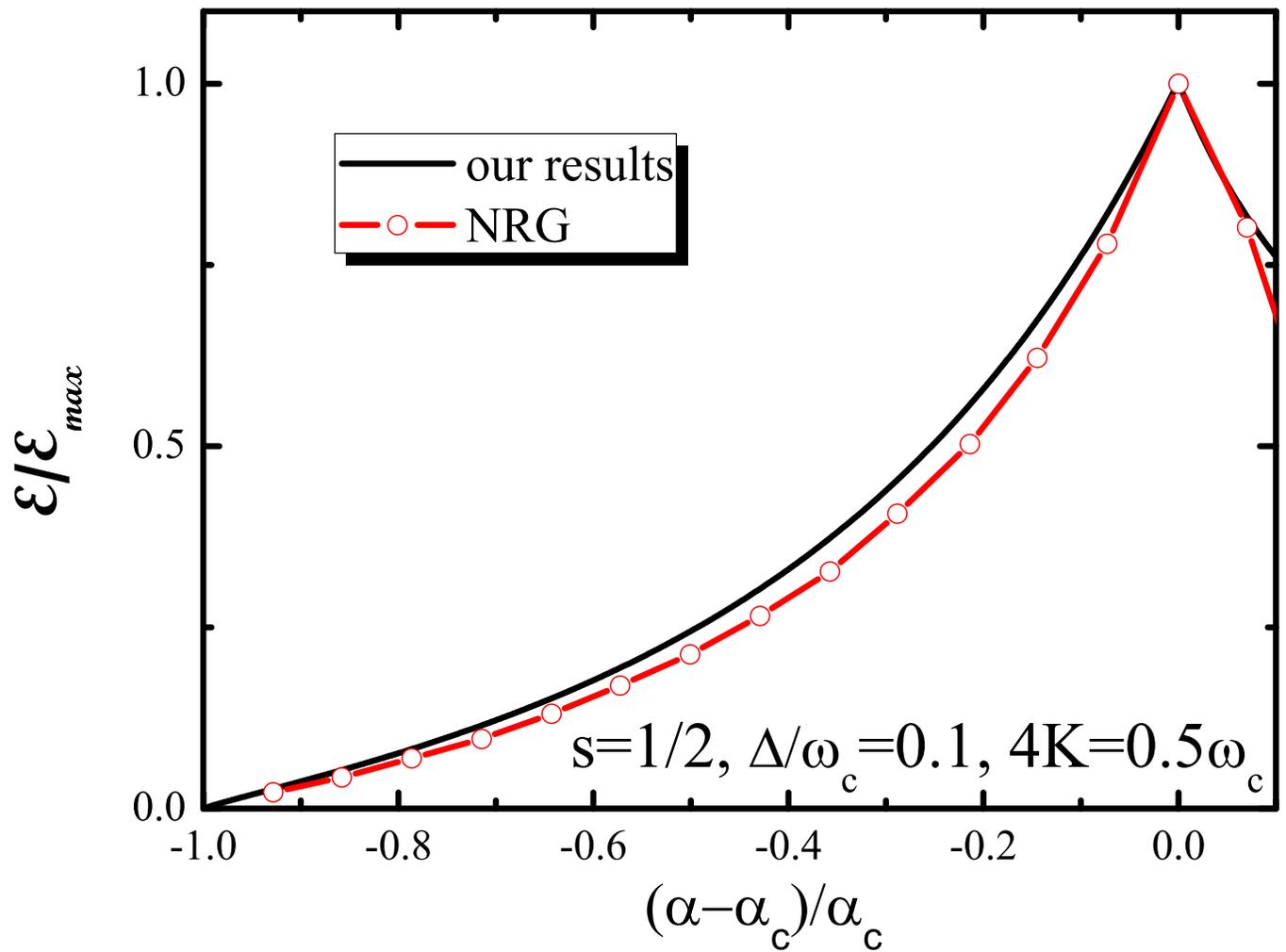
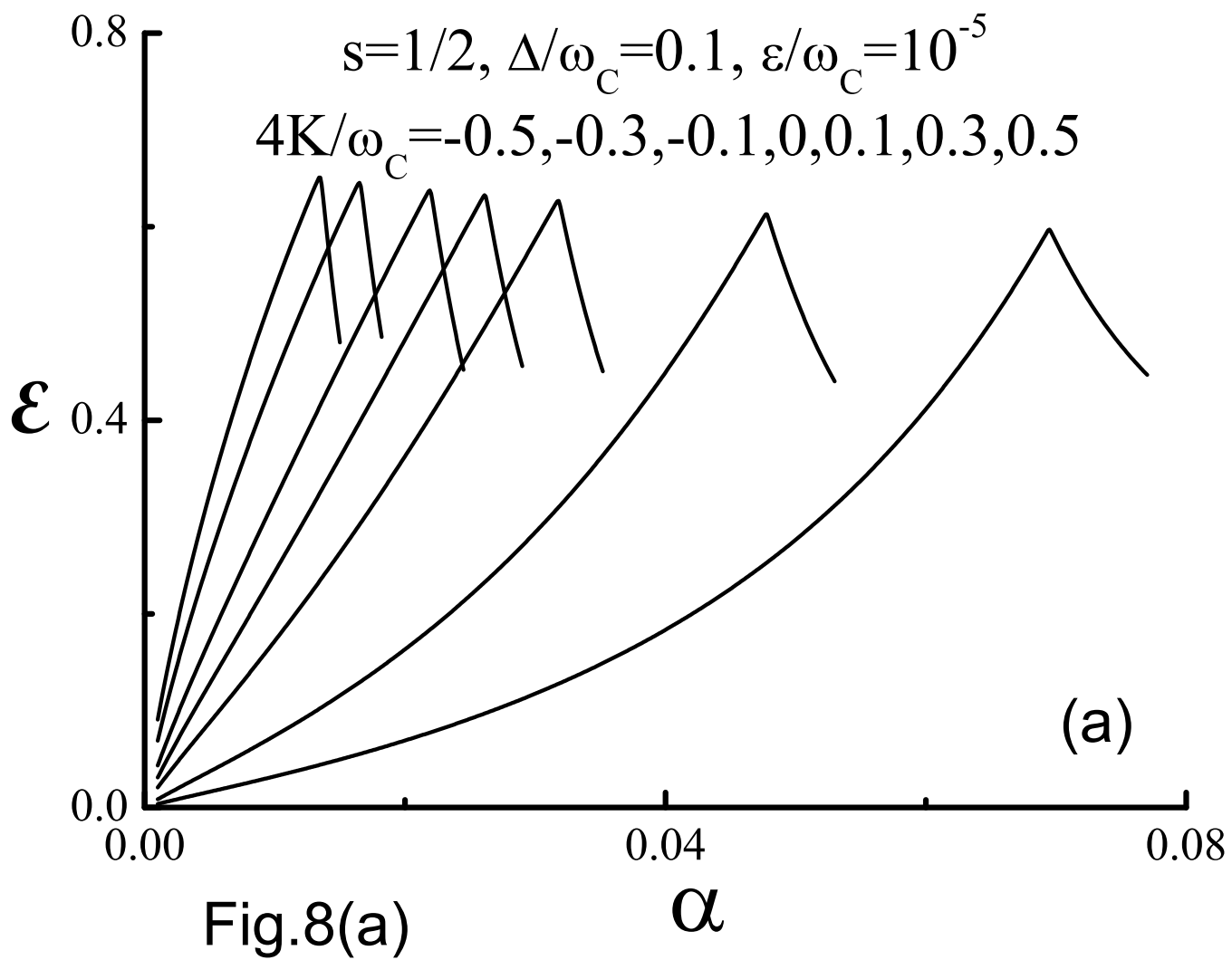
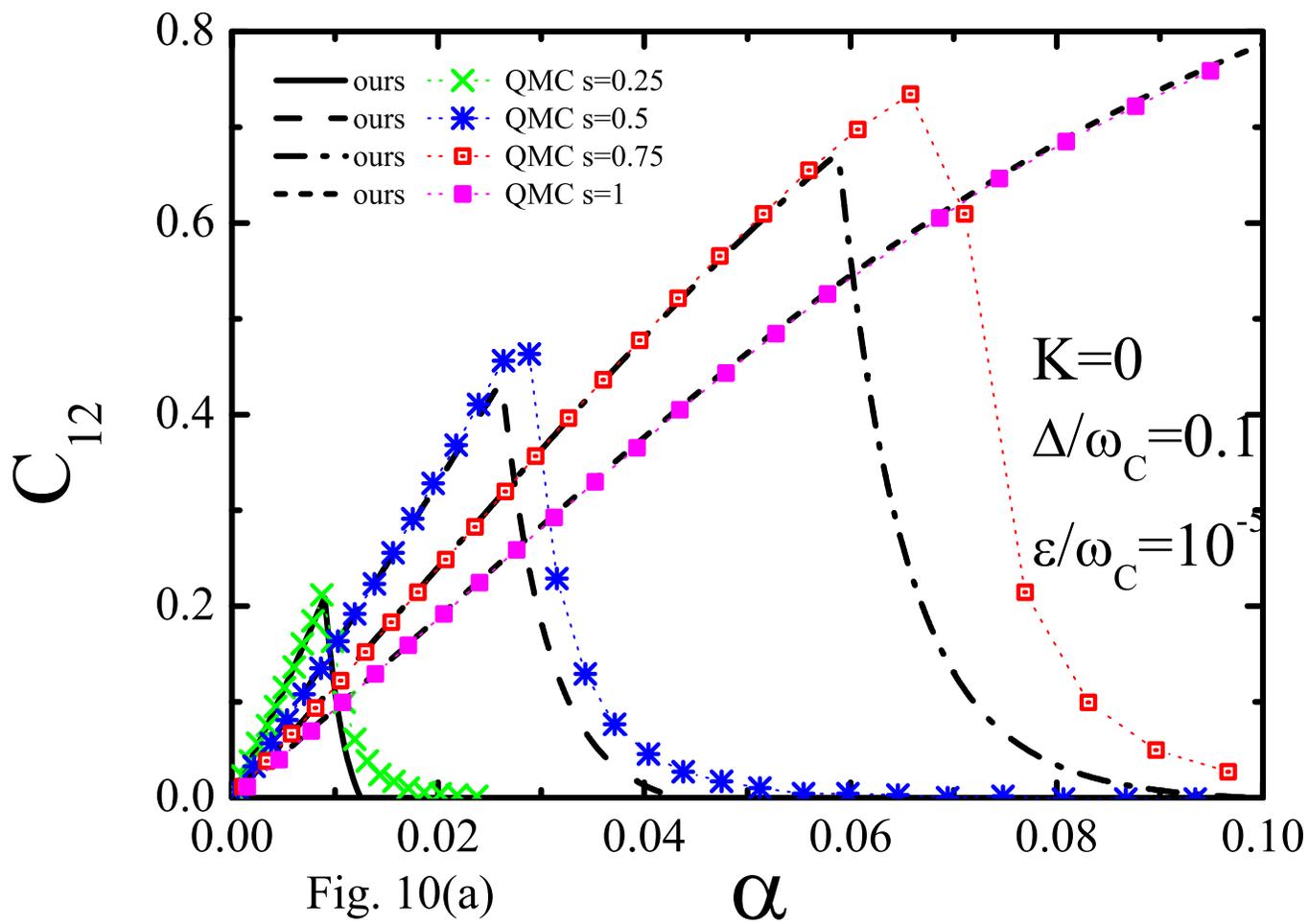
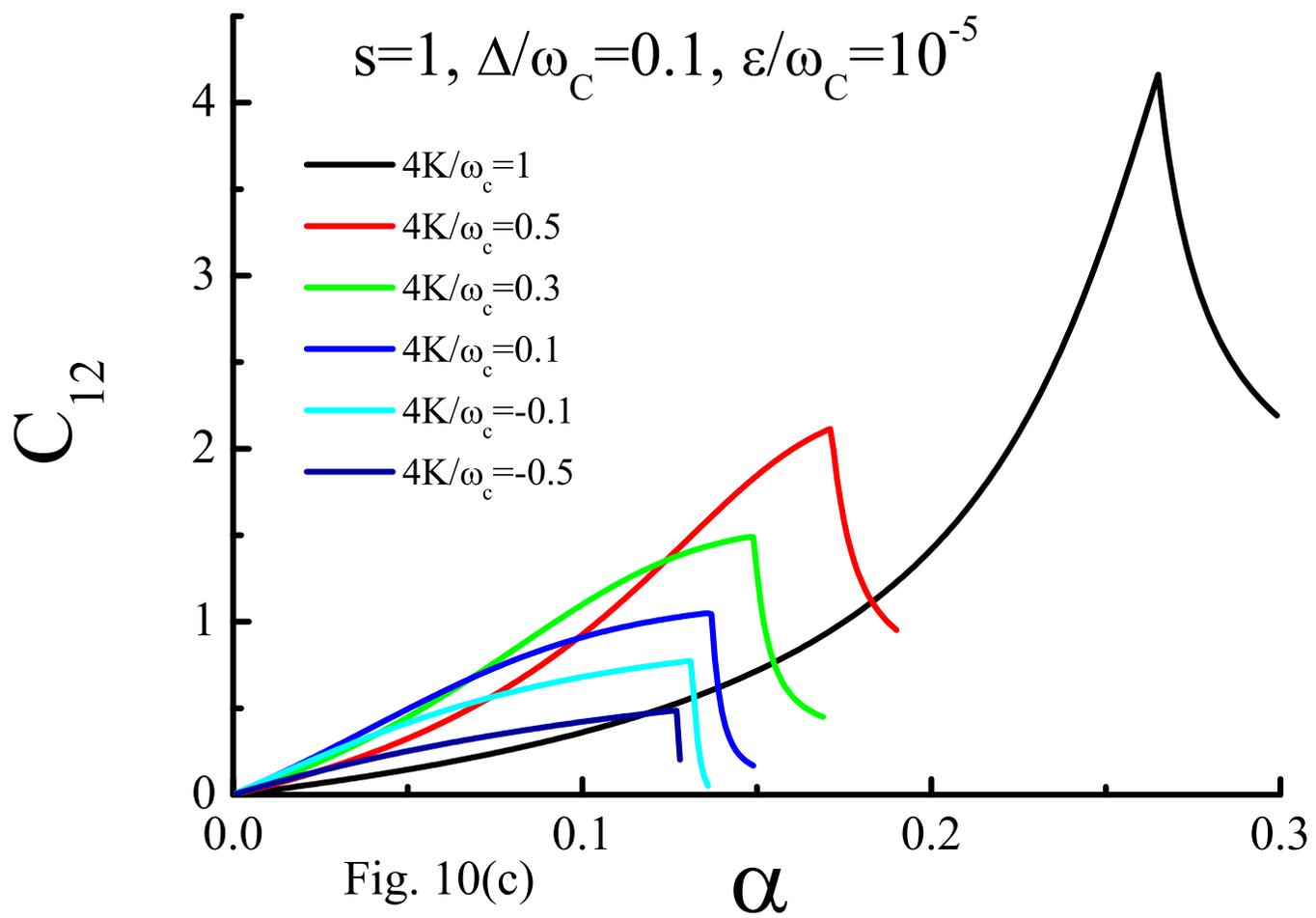
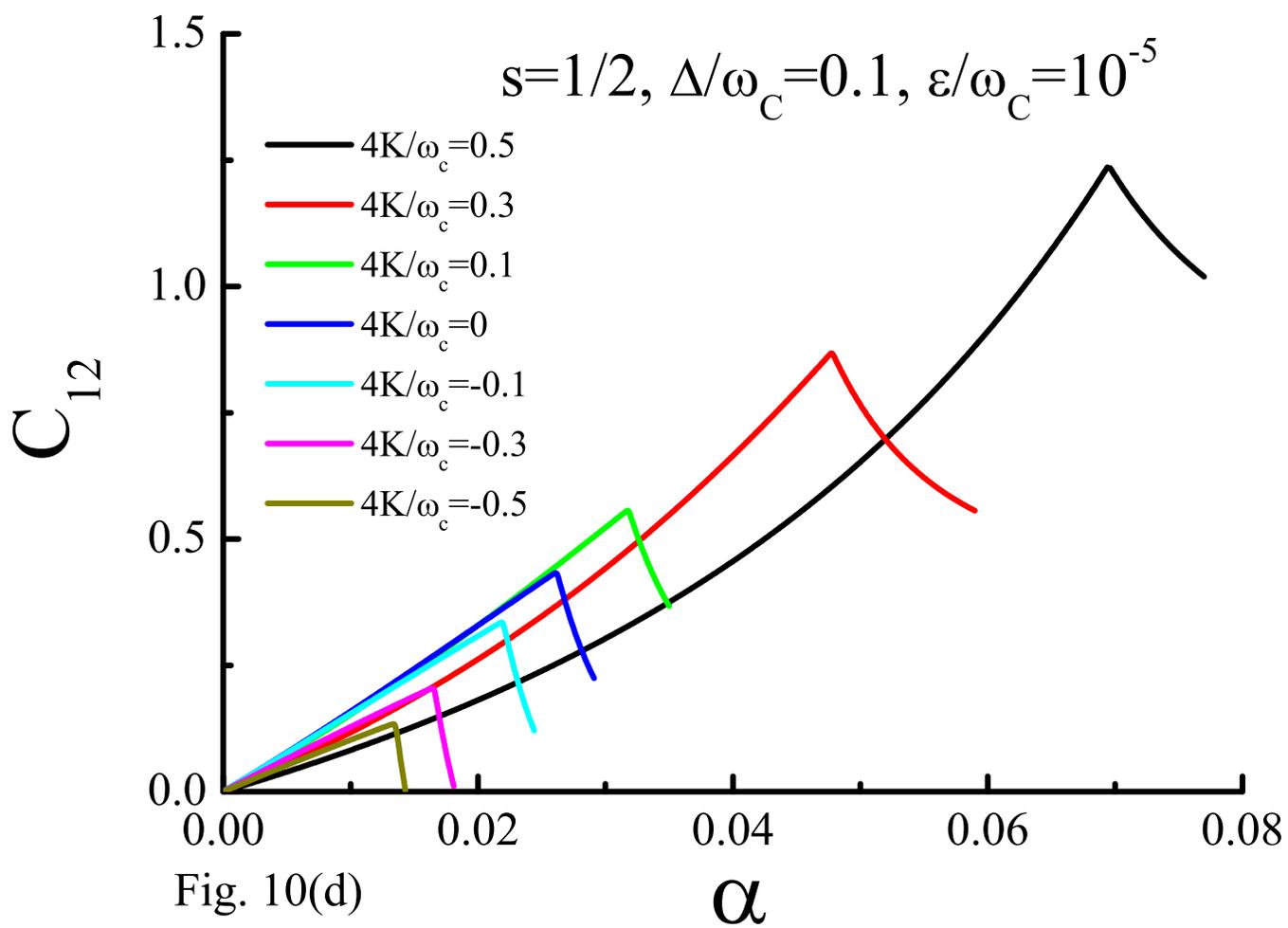
\end{document}